\documentclass[journal=cmatex,manuscript=article]{achemso}

\usepackage[version=3]{mhchem} % Formula subscripts using \ce{}
\usepackage{pslatex}
\usepackage{xr}
\usepackage{natbib}
\usepackage[dvipsnames]{xcolor} %Used to highlight parts of the text. Remove when submitting to journal (colorbox)

\DeclareGraphicsExtensions{.pdf,.png,.jpg}
\graphicspath{{./Figures/}} % this places all graphics in the figures subdirectory

\usepackage{tabularx,multirow} 
\usepackage{longtable} % to split a table across multiple pages
\usepackage{booktabs} %for the horizontal lines in a table %for \toprule, \midrule, \bottomrule

\author{Christopher L. Rom}
\affiliation[National Renewable Energy Laboratory]
{Materials Science Center, National Renewable Energy Laboratory, Golden, CO, 80401, USA}
\email{christopher.rom@nrel.gov}
\author{Matthew Jankousky}
\affiliation[Colorado School of Mines]
{Department of Metallurgical and Materials Engineering, Colorado School of Mines, Golden, Colorado 80401, USA}
\author{Maxwell Q. Phan}
\affiliation[National Renewable Energy Laboratory]
{Materials Science Center, National Renewable Energy Laboratory, Golden, CO, 80401, USA}
\author{Shaun O'Donnell}
\affiliation[National Renewable Energy Laboratory]
{Materials Science Center, National Renewable Energy Laboratory, Golden, CO, 80401, USA}
\author{Corlyn Regier}
\affiliation[Colorado State University]
{Department of Chemistry, Colorado State University, Fort Collins, CO, 80523-1872, USA}
\author{James R. Neilson}
\affiliation[Colorado State University]
{Department of Chemistry, Colorado State University, Fort Collins, CO, 80523-1872, USA}
\author{Vladan Stevanović}
\affiliation[Colorado School of Mines]
{Department of Metallurgical and Materials Engineering, Colorado School of Mines, Golden, Colorado 80401, USA}
\alsoaffiliation[National Renewable Energy Laboratory]
{Materials Science Center, National Renewable Energy Laboratory, Golden, CO, 80401, USA}
\author{Andriy Zakutayev}
\affiliation[National Renewable Energy Laboratory]
{Materials Science Center, National Renewable Energy Laboratory, Golden, CO, 80401, USA}
\email{Andriy.Zakutayev@nrel.gov}

\title{Ion exchange synthesizes layered polymorphs of \ce{MgZrN2} and \ce{MgHfN2}, two metastable semiconductors}

\begin{document}
\begin{abstract}
The synthesis of ternary nitrides is uniquely difficult, in large part because elemental \ce{N2} is relatively inert. 
However, lithium reacts readily with other metals and \ce{N2}, making Li-$M$-N the most numerous sub-set of ternary nitrides.
Here, we use \ce{Li2ZrN2}, a ternary with a simple synthesis recipe, as a precursor for ion exchange reactions towards \ce{$A$ZrN2} ($A$ = Mg, Fe, Cu, Zn). 
\textit{In situ} synchrotron powder X-ray diffraction studies show that \ce{Li+} and \ce{Mg^{2+}} undergo ion exchange topochemically, preserving the layers of octahedral [\ce{ZrN6}] to yield a metastable layered polymorph of \ce{MgZrN2} (spacegroup $R\overline{3}m$) rather than the calculated ground state structure ($I4_1/amd$). 
UV-vis measurements show an optical absorption onset near 2.0~eV, consistent with the calculated bandgap for this polymorph.
Our experimental attempts to extend this ion exchange method towards \ce{FeZrN2}, \ce{CuZrN2}, and \ce{ZnZrN2} resulted in decomposition products (\ce{$A$ + ZrN + 1/6 N2}), an outcome that our computational results explain via the higher metastability of these phases.
We successfully extended this ion exchange method to other Li-$M$-N precursors by synthesizing \ce{MgHfN2} from \ce{Li2HfN2}.
In addition to the discovery of metastable $R\overline{3}m$ \ce{MgZrN2} and \ce{MgHfN2}, this work highlights the potential of the 63 unique Li-$M$-N phases as precursors to synthesize new ternary nitrides. 

\end{abstract}

\section{Introduction}
Developing new methods for materials synthesis accelerates the rate at which new materials can be discovered. 
Prof. Francis DiSalvo epitomized this fact, as his development of ammonolysis and Na-flux synthesis methods unlocked many new ternary nitrides.\cite{marchand1999newRoutes, niewa1998recent, elder1993thermodynamics, yamane1997preparationNaFlux}
Yet ternary nitrides remain underexplored, with reported compounds lagging the number of known ternary oxides by an order of magnitude.\cite{greenaway2021ternaryReview, zakutayev2016designPhotovoltaicNitrides}
Thus, new methods are needed to continue charting this vast phase space.

Recent exploratory syntheses via metathesis reactions or thin film sputtering have yielded new ternary nitrides,\cite{zakutayev2022experimentalSynthesisNitrides} but they often form cation-disordered structures that deviate from the cation-ordered phases predicted by computational studies.\cite{sun2019map}
For example, \ce{MgZrN2} is predicted to form in the $\gamma$-\ce{LiFeO2} structure type ($I4_1/amd$), a cation-ordered variant of the rocksalt structure.
However, thin film\cite{bauers2019compositionMgZrN2, bauers2019ternaryrocksaltsemiconductors, bauers2020epitaxialMgZrN2} and bulk metathesis\cite{rom2021bulk, todd2021twostep} syntheses yielded the cation-disordered rocksalt structure ($Fm\overline{3}m$).
Similarly, \ce{ZnZrN2} has only been produced via sputtering that yielded cation-disordered rocksalt, h-BN, or antibixbyite structures (depending on cation ratios) even though the predicted ground state is a cation-ordered `wurtsalt' structure (with layers of tetrahedral wurtzite-like Zn and octahedral rocksalt-like Zr).\cite{woods2022roleZnZrN2}
In some cases, rapid thermal annealing can convert thin films with metastable disordered structures to stable ordered structures (e.g., \ce{MgMoN2}, \ce{MgWN2}, \ce{ScTaN2}, \ce{MgTa2N3}),\cite{zakutayev2024synthesis, rom2023bulkMgWN2} but synthesizing metastable ordered structures remains a challenge.

Chemical composition and structure are intimately linked to physical properties. Therefore, improved structural control for ternary nitrides will enable improved property control. As the disordered structures of the ternary nitrides described above differ from the predicted structures,\cite{sun2019map} this results in disagreement between the predicted and observed properties of these materials.\cite{bauers2019ternaryrocksaltsemiconductors} 
For example, the cation disorder commonly observed in sputtered thin films tends to decrease bandgaps.\cite{schnepf2020utilizing}
In the case of \ce{MgZrN2}, the ordered variant ($I4_1/amd$) is predicted to have an optical bandgap of 2.5~eV, but the disordered $Fm\overline{3}m$ variant that was experimentally synthesized showed an optical absorption onset of 1.8~eV.\cite{bauers2019ternaryrocksaltsemiconductors}
Improving our ability to synthesize cation-ordered polymorphs will therefore improve our ability to precisely realize computationally-predicted materials with desired properties.

One strategy for synthesizing new cation-ordered ternary nitrides is via ion exchange reactions. In these reactions, precursor cation-ordered ternaries serve as a template whereby the structure of the precursor is largely preserved in the ion-exchanged product. 
Previously, monovalent ions have been swapped to yield new delafossite phases (e.g., \ce{NaTaN2 + CuI -> CuTaN2 + NaI})\cite{zakutayev2014experimentalCuNbN2, yang2013strongCuTaN2, miura2011silver} and divalent ions have been exchanged to make new nitridosilicates (e.g., \ce{Sr2Si5N8 + 2CaCl2 -> $\beta$-Ca2Si5N8 + 2SrCl2})\cite{bielec2017increased, bielec2018fe2si5n8}
More recently, we have shown that two equivalents of a monovalent cation (\ce{Li+}) can be exchanged with one equivalent of a divalent cation (\ce{Zn^{2+}}) to synthesize \ce{Zn3WN4} via \ce{Li6WN4 + 3ZnBr2 -> Zn3WN4 + 6 LiBr}.\cite{rom2024low}
That work suggests Li-$M$-N precursors may hold promise for further heterovalent cation exchange reactions.

Here we report the synthesis of a metastable, cation-ordered polymorph of \ce{Mg$M$N2} ($M$ = Zr, Hf) in the \ce{\alpha-NaFeO2} structure type. 
This polymorph forms via a topochemical ion exchange reaction between \ce{Li2$M$N2} and \ce{Mg$X$2} ($X$ = Cl, Br). 
The layered configuration of the \ce{Li2ZrN2} precursor is retained through the reaction, as observed by \textit{in situ} synchrotron powder X-ray diffraction. 
This layered polymorph of \ce{MgZrN2} exhibits an optical absorption onset of 2.0 eV, consistent with calculations and slightly higher than the rocksalt polymorph (1.8 eV).
Our synthetic attempts towards \ce{FeZrN2}, \ce{CuZrN2}, and \ce{ZnZrN2} instead yielded decomposition products (\ce{$A$ + ZrN + 1/6 N2}), which we rationalize via DFT calculations that show these phases are more deeply metastable than \ce{MgZrN2}.
This work shows that Li-$M$-N ternaries are viable precursors for ion exchange syntheses using the examples of \ce{MgZrN2} and \ce{MgHfN2}, and advances synthesis science by exploring the limits of ion exchange reactions in ternary nitrides,

\section{Results and Discussion}
\subsection{Synthesis and structure of layered \ce{MgZrN2}}
\begin{figure}[ht!]
    \centering
    \includegraphics[width = 0.7\textwidth]{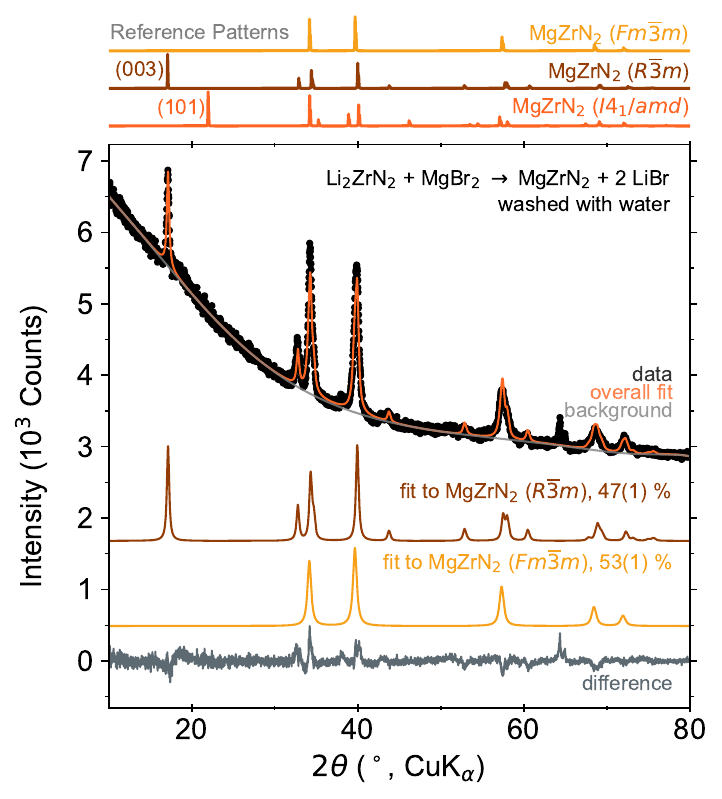}
    \caption{
    PXRD patterns of \ce{MgZrN2} synthesized from a reaction between \ce{Li2ZrN2 + MgBr2}, heated at 650~°C for 10 min.
    The LiBr byproduct has been washed away with water. 
    Simulated patterns for \ce{MgZrN2} polymorphs are shown for reference: $I4_1/amd$ (mp-1245429), $R\overline{3}m$ and $Fm\overline{3}m$ (this work).
    Peaks of an unknown impurity phase near 64° are not included in the fit.
    }
    \label{fig:xrd_MgZrN2_main}
    %Source: http://127.0.0.1:8888/notebooks/Documents/02_Nitrides/PXRD/PXRD%20plots%20for%20Li2MN2%20ion%20exchange.ipynb
    % CLR2_89A-2 washed
\end{figure}

\begin{figure}[ht!]
    \centering
    \includegraphics[width = \textwidth]{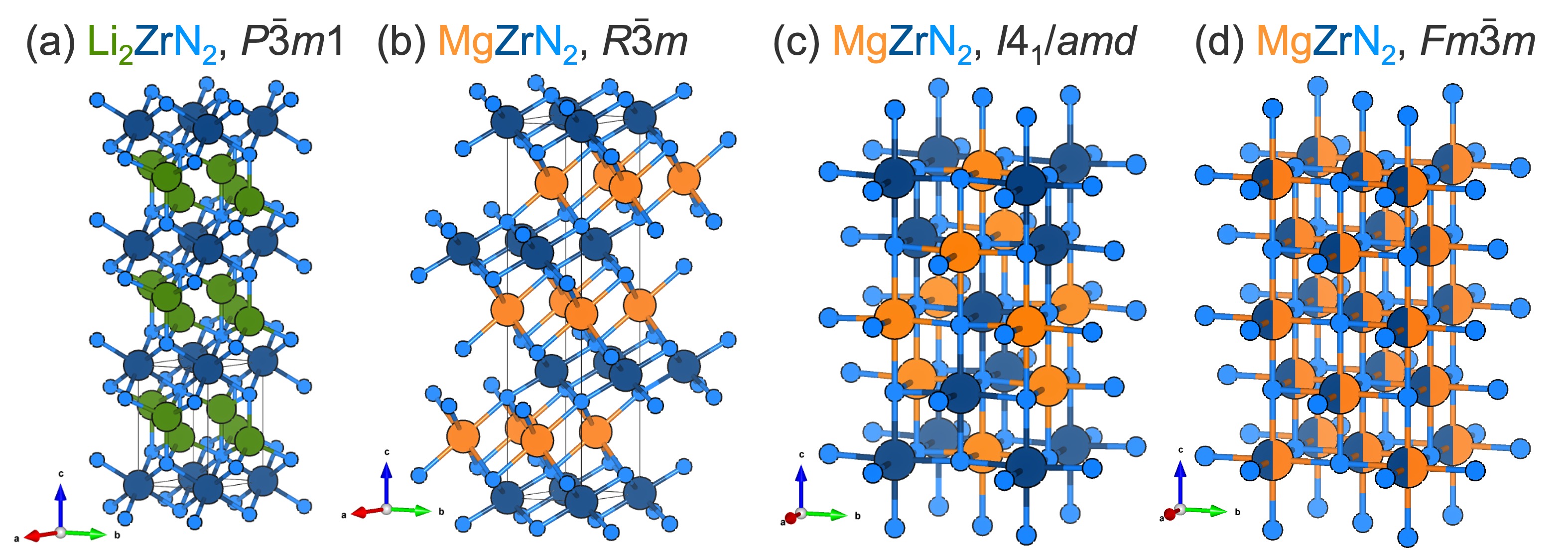}
    \caption{Crystal structures for (a) \ce{Li2ZrN2} ($P\overline{3}m1$ spacegroup), (b) metastable \ce{MgZrN2} in the $\alpha$-\ce{NaFeO2} structure type ($R\overline{3}m$), (c) stable \ce{MgZrN2} in the $\gamma$-\ce{LiFeO2} structure type ($I4_1/amd$), and (d) the cation-disordered rocksalt polymorph ($Fm\overline{3}m$) reported in prior work.\cite{bauers2019ternaryrocksaltsemiconductors, bauers2019compositionMgZrN2, bauers2020epitaxialMgZrN2, rom2021bulk, todd2021twostep}}
    \label{fig:structures_main}
\end{figure}

Reactions between \ce{Li2ZrN2 + MgBr2} produce \ce{MgZrN2} in the $\alpha$-\ce{NaFeO2} structure type (space group $R\overline{3}m$), along with a LiBr byproduct (Figure \ref{fig:xrd_MgZrN2_main}).
The LiBr can be washed away with water without changing the \ce{MgZrN2} structures (Figure S1). 
Powder X-ray diffraction (PXRD) prominently shows the (003) reflection of the $R\overline{3}m$ polymorph, whereas the primary reflection (101) of the $I4_1/amd$ polymorph does not appear (Figure \ref{fig:xrd_MgZrN2_main}). This indicates that the [\ce{ZrN6}] layers of the \ce{Li2ZrN2} precursor (Figure \ref{fig:structures_main}a) are retained in the $R\overline{3}m$ \ce{MgZrN2} product (Figure \ref{fig:structures_main}b), rather than undergoing a structural rearrangement to the $\gamma$-\ce{LiFeO2} form of cation ordering (Figure \ref{fig:structures_main}c).
However, the [\ce{ZrN6}] layers shift by x+2/3 and y+1/3 during the reaction (Figure S2). This shift changes the anion sublattice from hexagonal-close-packed (\ce{Li2ZrN2}) to cubic-close-packed (\ce{MgZrN2}). The cation sublattice is cubic-close-packed for both phases.

Rietveld analysis shows that the $Fm\overline{3}m$ polymorph of \ce{MgZrN2} (a cation disordered rocksalt structure) is also present (Figure \ref{fig:xrd_MgZrN2_main}).
While the $R\overline{3}m$ phase accurately describes all the peak positions, including the $Fm\overline{3}m$ secondary phase better captures the peak intensities and improves the fit quality from $R_\mathrm{wp} = 5.749$\% (with $R\overline{3}m$ alone) to 4.336\% (with both the $R\overline{3}m$ and $Fm\overline{3}m$ structures).
We observed similar partial disorder for wurtzite-derived \ce{Zn3WN4} synthesized via a similar ion exchange process.\cite{rom2024low}
The lattice parameter of this $Fm\overline{3}m$ \ce{MgZrN2} matches prior reports on this phase ($a = 4.54$~Å),\cite{rom2021bulk, bauers2019compositionMgZrN2} and is smaller than the parameter for rocksalt ZrN ($a = 4.58$~Å).\cite{christensen_neutron_1975}
However, the phase may exhibit some degree of Mg loss or oxide incorporation (e.g., \ce{Mg$_{1-x}$Zr$_{1+x}$N$_{2-y}$O$_y$}) that cannot be accurately refined from these XRD data. 
Such off-stoichiometry is common for rocksalt \ce{MgZrN2}.\cite{bauers2019compositionMgZrN2, rom2021bulk, kim2020influenceMgZrN2}
Heating to higher temperatures or for longer times decreases the relative intensity of the (003) reflection, and increases the fraction of disordered $Fm\overline{3}m$ \ce{MgZrN2} relative to the ordered $R\overline{3}m$ polymorph (Figure S3, Table S1).

The layered $R\overline{3}m$ structure of \ce{MgZrN2} ($\alpha$-\ce{NaFeO2}-type) that we report here is different from prior experimental and computational reports on this composition. Thin film sputtering experiments\cite{bauers2019compositionMgZrN2, bauers2019ternaryrocksaltsemiconductors, sun2019map, zakutayev2022experimentalSynthesisNitrides, kim2020influenceMgZrN2, bauers2020epitaxialMgZrN2} and bulk solid state metathesis reactions\cite{rom2021bulk, todd2021twostep} both produced \ce{MgZrN2} exclusively in the disordered rocksalt structure ($Fm\overline{3}m$), along with related solid solutions (\ce{Mg_{$x$}Zr_{2-$x$}N2}; $0 \leq x \leq 1$). 
In contrast, some computational studies have identified a $I4_1/amd$ polymorph ($\gamma$-\ce{LiFeO2}-type) as the thermodynamic ground state,\cite{xue2021electronicMgZrN2, arslan2024investigatingDFT_MgZrN2} while one group has argued that the $\gamma$-\ce{LiFeO2} ordering (relaxed into a monoclinc \ce{LiUN2}-type structure) and $\alpha$-\ce{NaFeO2} ordering are nearly equal in energy.\cite{gharavi2021theoretical} 
Both the $I4_1/amd$ ($\gamma$-\ce{LiFeO2}) and $R\overline{3}m$ ($\alpha$-\ce{NaFeO2}) structures are cation-ordered variants of the rocksalt structure (Figure \ref{fig:structures_main}).\cite{mather2000review} 
Our calculations find that the $R\overline{3}m$ polymorph synthesized here is +42~meV/formula unit higher in energy compared to the ground state $I4_1/amd$ polymorph.
The fact that longer heating times decreases the proportion of the $R\overline{3}m$ polymorph relative to the $Fm\overline{3}m$ polymorph supports the computational results that $R\overline{3}m$ is metastable (Table S1).

\subsection{Optical properties}
\begin{figure}[ht!]
    \centering
    \includegraphics{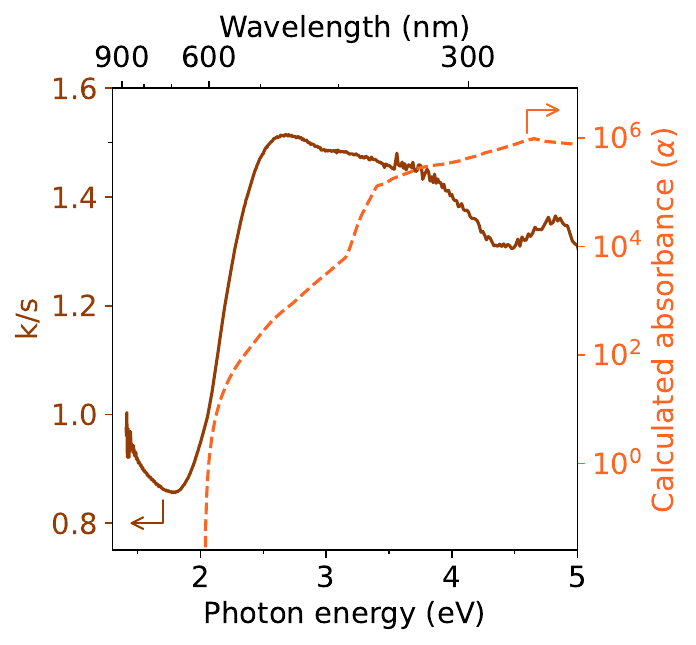}
    \caption{UV-vis diffuse-reflectancee spectroscopy measurement for $R\overline{3}m$ \ce{MgZrN2} (solid trace) compared with the calculated absorption spectrum (dashed trace).}
    \label{fig:uv_vis_main}
    % Source: http://localhost:8888/notebooks/Documents/02_Nitrides/UVvis/UV_vis_MgZrN2.ipynb
\end{figure}

Optical measurements of the un-optimized $R\overline{3}m$ \ce{MgZrN2} synthesized here are consistent with the bandgap predicted in prior literature (Figure \ref{fig:uv_vis_main}). 
The Kubelka-Munk transform of UV-vis diffuse-reflectance spectroscopy on $R\overline{3}m$ \ce{MgZrN2} shows an optical absorption onset of 2.0~eV. 
Similarly, GW calculations show an indirect bandgap of 1.95~eV for the $R\overline{3}m$ polymorph (NREL MatDB ID \#290018),\cite{lany2013band, lany2015semiconducting} with a direct transition just above 2.0~eV (dashed trace).
In contrast, the $I4_1/amd$ structure is calculated to have an electronic bandgap of 1.47~eV (NREL MatDB ID \#290029) with a direct-but-forbidden transition and an optical bandgap of 2.5~eV.\cite{bauers2019ternaryrocksaltsemiconductors}
Thin films of $Fm\overline{3}m$ \ce{MgZrN2} (disordered rocksalt) show an absorption onset at 1.8 eV,\cite{bauers2019ternaryrocksaltsemiconductors, bauers2019compositionMgZrN2} while $Fm\overline{3}m$  \ce{MgZrN2} made via bulk metathesis reactions did not show any clear absorption onset.\cite{rom2021bulk}
Longer annealing times for this synthesis led to a darker powder with a weaker absorption onset (Figure S4). 
The increased absorbance likely stems from a metallic (or highly defective) phase in the rocksalt structure (e.g., a magnesium-poor \ce{Mg_{1-x}Zr_{1+x}N2}, a nitrogen-poor \ce{MgZrN_{2-\delta}}, or an oxynitride \ce{MgZrN_{2-\delta}O_{\delta}}), as the rocksalt phase fraction increased with the darkening of the powder (Table S1).
The clear absorption onset for $R\overline{3}m$ \ce{MgZrN2} suggests this new material may have promising optoelectronic properties for visible light absorption, but further work will be necessary to  optimize the synthesis of this material. 

\subsection{\textit{In situ} synchrotron X-ray powder diffraction}
\begin{figure}[ht!]
    \centering
    \includegraphics{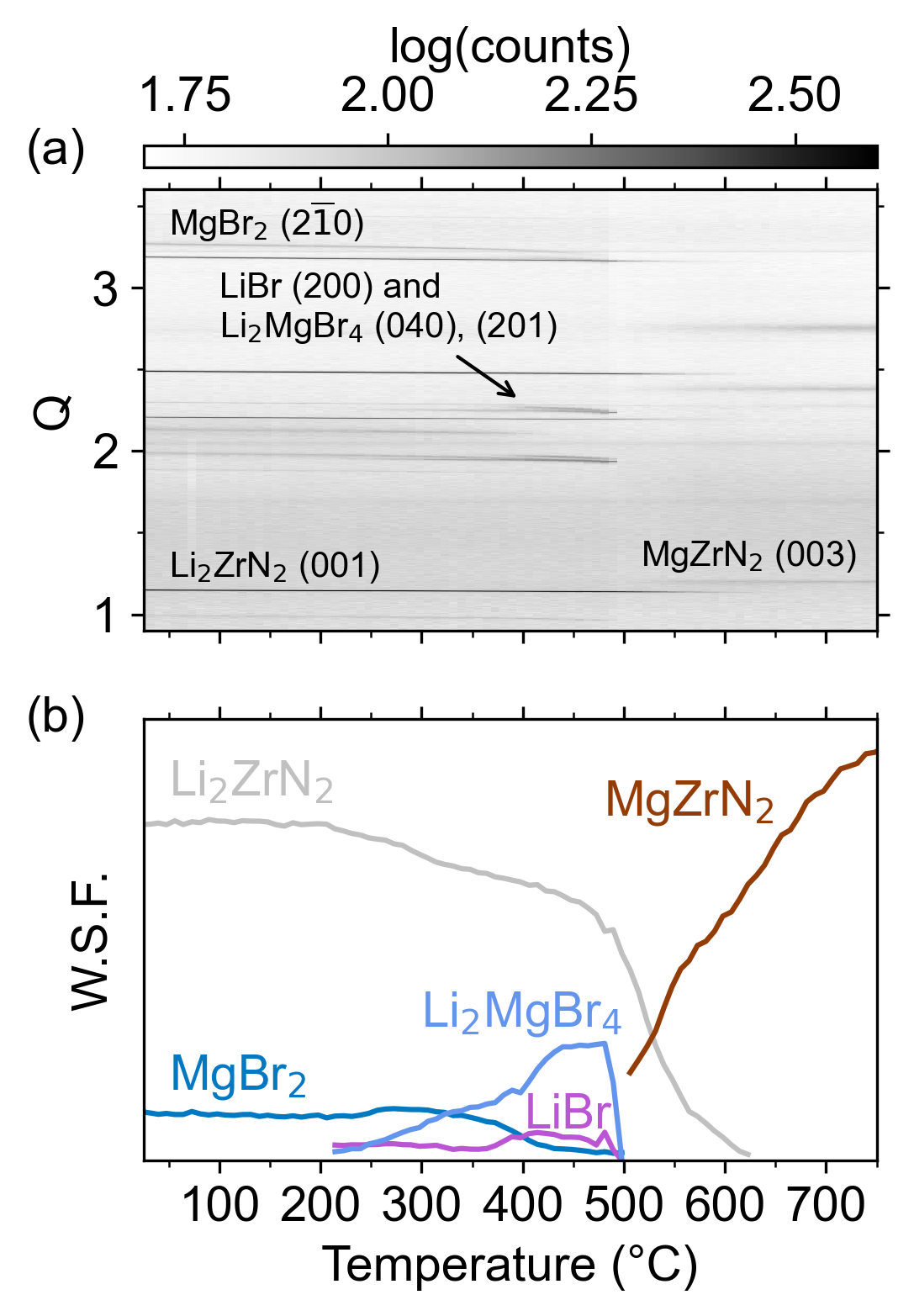}
    \caption{(a) \textit{In situ} SPXRD measurements of \ce{Li2ZrN2 + MgBr2} as a function of temperature, with select Bragg reflections labeled. (b) Sequential Rietveld analysis of the SPXRD data shows the relative amounts of crystalline material, expressed as a weighted scale factor (W.S.F.). Select fits are shown in Figure S5.}
    \label{fig:inSitu_main}
    % Source: http://localhost:8888/notebooks/OneDrive%20-%20NREL/Beamtime_202404_refinements/workup_MgZrN2_MgHfN2.ipynb 
\end{figure}
\textit{In situ} synchrotron powder X-ray diffraction (SPXRD) shows that \ce{Li2ZrN2} undergoes ion exchange with \ce{MgBr2} at moderate temperatures to produce cation-ordered \ce{MgZrN2} (Figures \ref{fig:inSitu_main} and S5). The SPXRD data were modeled using the Rietveld method to calculate weighted scale factors (W.S.F.) for each phase, which is a proxy for the amount of crystalline material present (see methods). Initially, \ce{Li2ZrN2} and \ce{MgBr2} are present, although the \ce{MgBr2} peaks are fairly weak, possibly owing to poor crystallinity of this precursor. As the temperature increases above 200~°C, reactivity begins, as evidenced by the W.S.F. decreasing for \ce{Li2ZrN2} and the growth of \ce{LiBr} and \ce{Li2MgBr4} phases. The \ce{Li2MgBr4} is likely an incidental intermediate formed as precursor and product salts react with one another: \ce{2 LiBr + MgBr2 -> Li2MgBr4}. Near 500~°C, the halide peaks disappear, indicating a melting process. At the same time, the \ce{MgZrN2} peaks grow rapidly and the decline of the \ce{Li2ZrN2} peaks accelerates. The reaction is complete by 640~°C, as the \ce{Li2ZrN2} peaks completely disappear. These diffraction data suggest that the layered structure of \ce{Li2ZrN2} (Figure \ref{fig:structures_main}a) templates the layered structure of $R\overline{3}m$ \ce{MgZrN2} (Figure \ref{fig:structures_main}b). 

\subsection{Thermochemical calculations and comparison with experiment}
The successful synthesis of a metastable, layered polymorph of \ce{MgZrN2} inspired us to further explore this ion exchange reaction in attempts to synthesize other layered \ce{$A$ZrN2} compounds ($A = $ Fe, Cu, Zn). DFT calculations show that such compounds are thermodynamically stable against the elements (Table \ref{tab:dft_and_insitu_summary}). However, \textit{ex situ} PXRD (Figures S9-S10) and \textit{in situ} SPXRD (Figures S6-S8) experiments show no evidence of \ce{FeZrN2}, \ce{CuZrN2}, or \ce{ZnZrN2}. Instead metallic Fe, Cu, and Zn are observed at low temperatures, indicating nitrogen loss from the solid: \ce{Li2ZrN2 + $AX_2$ -> $A$ + 1/2 N2 + ZrN + 2 Li$X$} (where $X$ = Cl, Br). These experiments do not rule out the possibility of synthesizing these phases, but they do suggest that other synthetic methods or conditions may be needed. Only \ce{MgZrN2} (and \ce{MgHfN2}, see below) exhibited sufficient stability under the temperatures required to drive the ion exchange reactions to completion (ca. 550-650°C).

\begin{table}[ht!]
\begin{tabular}{lllll}
Target & $\Delta H_f$ (eV/atom) & \textit{In situ} SPXRD data & Reactants       & Observed products \\ \hline
\ce{MgZrN2} & -1.58 %-2.057
& Figure \ref{fig:inSitu_main} & \ce{MgBr2 + Li2ZrN2} & \ce{MgZrN2}, LiBr      \\
\ce{FeZrN2} & -0.91 %-1.875
& Figure S6 & \ce{FeCl2 + Li2ZrN2} & Fe, ZrN, LiCl     \\
\ce{CuZrN2} & -0.60 %-1.429
& Figure S7 & \ce{CuBr2 + Li2ZrN2} & Cu, ZrN, LiBr     \\
\ce{ZnZrN2} & -0.98 %-1.449
& Figure S8 & \ce{ZnBr2 + Li2ZrN2}* & Zn, ZrN, LiBr     \\ \hline
\end{tabular}
\caption{Targeted compositions and DFT computed formation enthalpies with FERE corrections to the chemical potentials of elements for \ce{$A$ZrN2} phases in the $\alpha$-\ce{NaFeO2} structure type, along with the summary of \textit{in situ} SPXRD reactant mixtures and observed products. *A LiCl/KCl flux was added to the Zn-based reaction to act as a heat sink.}
\label{tab:dft_and_insitu_summary}
\end{table}

\begin{figure}[ht!]
    \centering
    \includegraphics{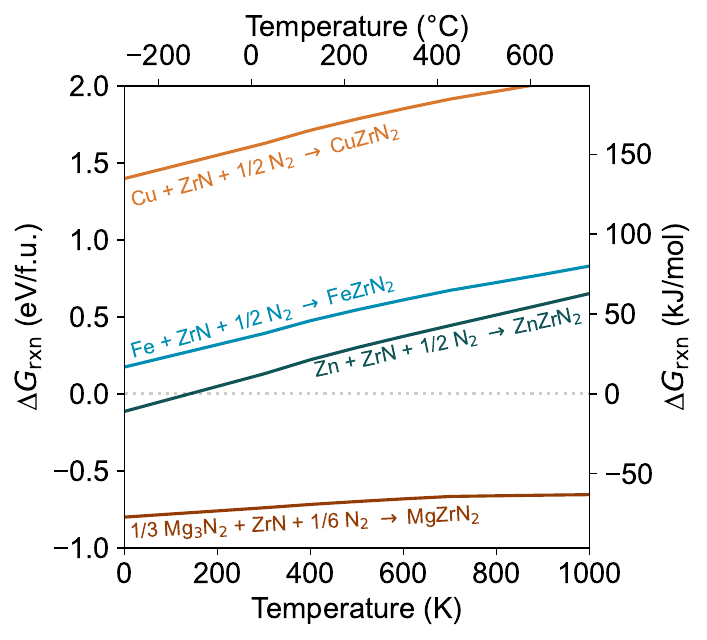}
    \caption{Reaction energies as a function of temperature for \ce{$A$ZrN2} phases in the $\alpha$-\ce{NaFeO2} structure type, compared with elemental or binary competing phases. Positive values indicate thermodynamic instability for the \ce{$A$ZrN2} phase.}
    \label{fig:dG_T}
    % source: http://localhost:8888/notebooks/Documents/02_Nitrides/manuscript_layered_MgZrN2/DFT_from_Matt/dG_calcs_for_AZrN2_stability.ipynb 
\end{figure}

To better understand the thermodynamic (in)stability of these layered nitrides, we calculated $\Delta G_\mathrm{rxn}$ as a function of temperature using a machine-learning-derived approximation (Figure \ref{fig:dG_T}).\cite{bartel_physical_2018} 
As the overall reactions energies are substantially negative for the \ce{Li2ZrN2 + ZnBr2 -> ZnZrN2 + 2LiBr} and \ce{Li2ZrN2 + MgBr2 -> MgZrN2 + 2LiBr} reactions (Figure S11), the differing outcomes likely relate to the stability of the \ce{$A$ZrN2} phases rather than overall $\Delta G_\mathrm{rxn}$.
Temperature-dependent free energy calculations show \ce{MgZrN2} is stable against decomposition across the temperature range explored (Figure \ref{fig:dG_T}).
In contrast, $\alpha$-\ce{NaFeO2}-type \ce{FeZrN2}, \ce{CuZrN2}, and \ce{ZnZrN2} are thermodynamically unstable when compared to \ce{$A$ + ZrN + 1/2 N2} at finite temperatures. 
These calculations are consistent with the \textit{in situ} diffraction experiments, which show the formation of \ce{$A$ + ZrN} at low temperatures (Figures S6-S8).
We note that a `wurtsalt' structure of \ce{ZnZrN2} ($P3m1$, with layers of tetrahedrally-coordinated \ce{Zn^{2+}}) is calculated to be thermodynamically stable,\cite{woods2022roleZnZrN2} but we do not observe this phase either (Figure S8). In our synthesis of \ce{MgZrN2}, we note cation disorder and possible metal-poor conditions (Figure S3).
Zn-poor conditions and cation disorder destabilize the wurtsalt phase relative to the rocksalt structure,\cite{woods2022roleZnZrN2} 
and therefore, the disordered rocksalt phase of \ce{ZnZrN2} is likely the more relevant phase for these reaction conditions.
Additionally, vacuum conditions led to decomposition of the disordered \ce{ZnZrN2} thin film materials above 300 °C,\cite{woods2022roleZnZrN2} and our syntheses were conducted in evacuated ampules or capillaries.
However, work on another Zn-ternary nitride, \ce{Zn2NbN3}, showed that annealing films under flowing \ce{N2} prevented decomposition until $>550$°C.\cite{zakutayev2021synthesisZn2NbN3} 
This prior literature suggests that bulk synthesis routes to cation-ordered \ce{ZnZrN2} might still be possible, but more work must be done to identify the necessary precursors and reaction conditions to facilitate reactivity at lower temperature or higher pressure. 

Layered structures of \ce{FeZrN2} and \ce{ZnZrN2} may yet be synthetically accessible. 
Based on prior literature, ion exchange reactions are capable of synthesizing ternary nitrides that are up to +0.5~eV/f.u. metastable with respect to decomposition products. 
For example, metastable delafossite structures of \ce{Cu$M$N2} ($M$ = Nb, Ta) were synthesized via ion exchange reactions: \ce{CuI + Na$M$N2 -> Cu$M$N2 + NaI}. 
These phases exhibited high metastabilities:  $\Delta H_\mathrm{rxn} = +0.48$~eV/f.u. for \ce{Cu + 1/6 Nb6N5 + 7/6 N2 -> CuNbN2}, and $\Delta H_\mathrm{rxn} = +0.42$~eV/f.u. for \ce{Cu + 1/3 Ta3N5 + 1/6 N2 -> CuTaN2}.\cite{zakutayev2014experimentalCuNbN2, yang2013strongCuTaN2}
The use of iodide anions in these reactions may be essential, as the smaller $\Delta H_f$ for NaI (-287.8~kJ/mol) likely produces less self-heating (and decomposition) compared to the more exothermic reactions yielding NaBr ($\Delta H_f$ = -361.1 kJ/mol) or NaCl ($\Delta H_f$ = -411~kJ/mol).\cite{crcHandbook_Thermo}
This trend also holds true for exchanges between Fe/Cu/Zn halides and Li halides (Figure S12).
Similar ion exchange reactions in ternary oxides have accessed even more deeply metastable phases such as \ce{Sn(Zr_{1/2}Ti_{1/2})O3} (+0.5 eV/atom; ca. +2.5~eV/f.u. metastable compared to the binaries).\cite{odonnell2022prediction}

% \clearpage
\subsection{Generalizability of Li-$M$-N precursors}

\begin{figure}[ht!]
    \centering
    \includegraphics[width=\linewidth]{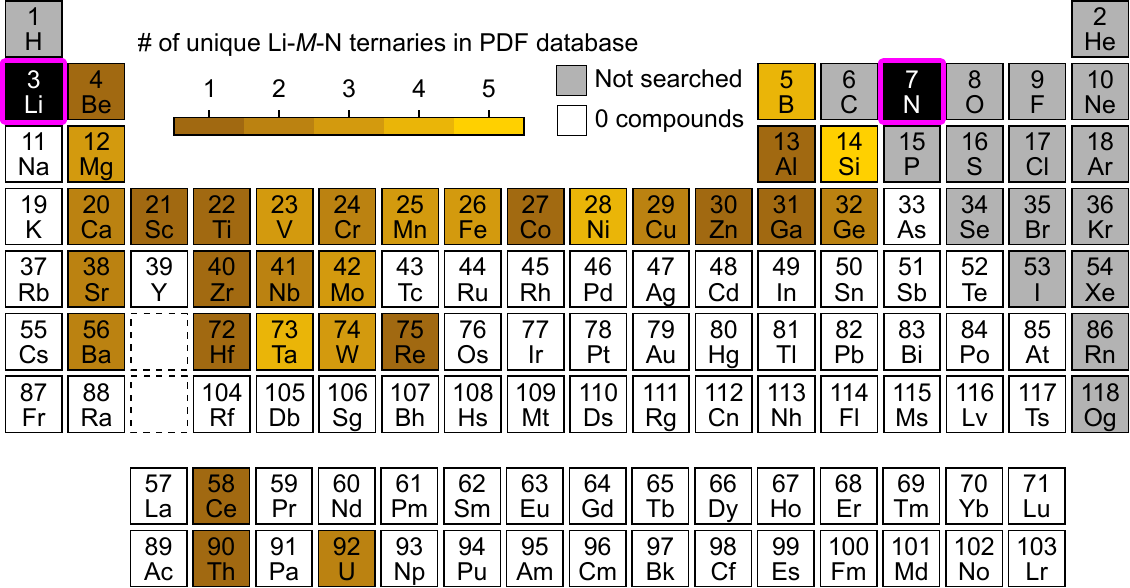}
    \caption{Periodic table of unique Li-$M$-N ternaries reported in the ICDD PDF database.}
    \label{fig:ptable}
    % Source: /Users/crom/Documents/02_Nitrides/ptable_Li-M-N
    % Searched on Friday 8/23/24 using the PDF database that's in the FTLB computer
\end{figure}

\begin{figure}[ht!]
    \centering
    \includegraphics[width = 0.7\textwidth]{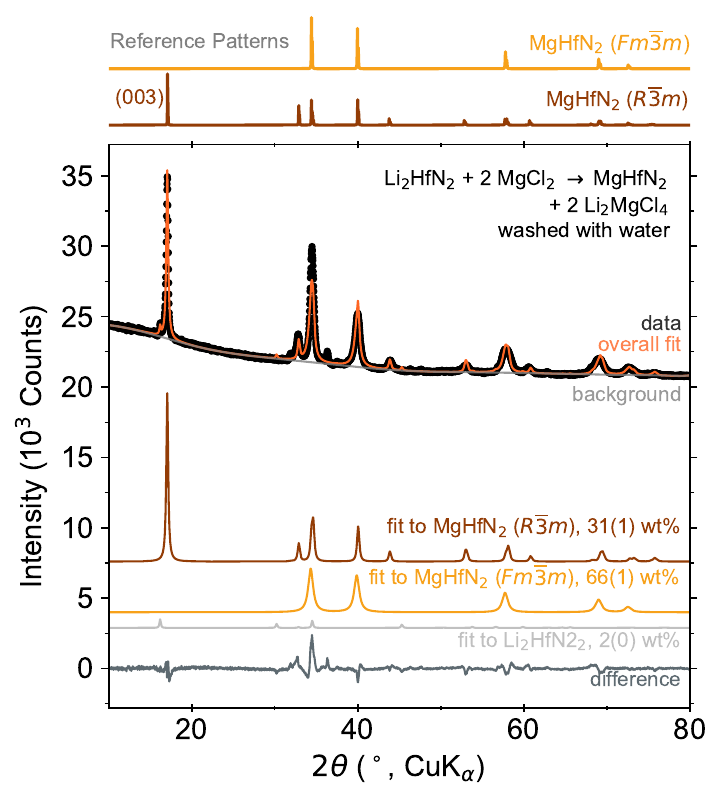}
    \caption{Powder X-ray diffraction patterns of \ce{MgHfN2} synthesized from a reaction between \ce{Li2HfN2 + 2 MgCl2}, heated at 550~°C for 10~h. % in a sealed, evacuated ampule. 
    The \ce{Li2MgCl4} byproduct has been washed away with water. Simulated patterns are shown for reference: \ce{MgHfN2} ($Fm\overline{3}m$) and \ce{MgHfN2} ($R\overline{3}m$).
    }
    \label{fig:xrd_MgHfN2_suppInfo}
%Source: http://127.0.0.1:8888/notebooks/Documents/02_Nitrides/PXRD/PXRD%20plots%20for%20Li2MN2%20ion%20exchange.ipynb
\end{figure}

Starting with Li-$M$-N precursors is a generalizable approach for synthesizing new ternary nitrides. 
Of the synthesized ternary nitrides, Li-$M$-N phases are the most well-studied subclass,\cite{greenaway2021ternaryReview} with at least 30 possible options for $M$ (Figure \ref{fig:ptable}, Tables S2 and S3). 
As a simple example, we synthesized $R\overline{3}m$ \ce{MgHfN2} via an ion exchange reaction (Figure \ref{fig:xrd_MgHfN2_suppInfo}). We used an excess of \ce{MgCl2} to drive the reaction \ce{Li2HfN2 + 2 MgCl2 -> MgHfN2 + Li2MgCl4}, and subsequently washed the sample with water to remove the halide (Figure \ref{fig:xrd_MgHfN2_suppInfo}). The (003) reflection for the \ce{MgHfN2} $R\overline{3}m$ structure is prominent. Including preferred orientation in the Rietveld analysis is required to properly fit this peak. The reaction was conducted at 550 °C for a 10~h dwell, which was insufficient to drive full conversion, as indicated by a small amount of residual \ce{Li2HfN2}. As \ce{Li2HfN2} is moisture sensitive, we hypothesize that this phase was protected during the washing procedure by a shell of \ce{MgHfN2}. This result demonstrates how these ion exchange reactions may be generalizable for synthesizing other $A$-$M$-N phases.

Some elements form multiple Li-$M$-N compounds with different compositions and structures, which may allow for the synthesis of various $A$-$M$-N structures. 
For example, W is known to form \ce{Li6WN4} and \ce{LiWN2}.\cite{yuan2005synthesisLi6WN4, kuhn2007effectLiWN2} 
The \ce{Li6WN4} has tetrahedrally coordinated \ce{WN4} units in an antifluorite structure,\cite{yuan2005synthesisLi6WN4} which we previously used to synthesize the wurtzite-like \ce{Zn3WN4} via a reaction between \ce{Li6WN4 + 3 ZnBr2}.\cite{rom2024low}
In contrast, \ce{LiWN2} is a layered compound with trigonal prismatic \ce{WN6} units,\cite{kuhn2007effectLiWN2} and it would be interesting to see if \ce{LiWN2} can undergo ion exchange to preserve those layers.
This example highlights how Li-$M$-N phases may be useful precursors for rapidly accelerating materials discovery for ternary nitrides. 

\section{Conclusion}
We report the synthesis of a new metastable polymorph of \ce{MgZrN2} and \ce{MgHfN2} in the $\alpha$-\ce{NaFeO2} structure type. Powder X-ray diffraction measurements confirmed that cation-ordered nature of this material, which contrasts with prior bulk and thin-film syntheses that produced a cation-disordered rocksalt structure. This material also differs from prior computational studies that predicted a $\gamma$-\ce{LiFeO2} structure type as the thermodynamic ground state. The synthesized \ce{MgZrN2} exhibits an optical absorption onset near 2.0~eV, consistent with calculated absorbance spectra for the metastable layered structure ($R\overline{3}m$). \textit{In situ} synchrotron powder X-ray diffraction measurements reveal the direct ion exchange pathway from the \ce{Li2ZrN2 + MgBr2} precursors to the \ce{MgZrN2 + 2 LiBr} products. The key feature of this synthesis is the layered nature of the \ce{Li2ZrN2} precursor, which templates the formation of the layered \ce{MgZrN2} structure. While first principles calculations (at T = 0~K) suggest that other layered \ce{$A$ZrN2} phases ($A$ = Fe, Cu, Zn) may be synthesizable, temperature-dependent $\Delta G$ calculations show that these phases become less stable with increasing temperature. Our \textit{in situ} X-ray diffraction measurements did not reveal these materials. However, we were able to successfully synthesize layered \ce{MgHfN2} via this ion exchange method, demonstrating how Li-$M$-N phases may be used as precursors to make other $A$-$M$-N compounds. These results show that ion exchange reactions are a promising strategy for improved structural control in the synthesis of emerging ternary nitride semiconductors.

\section{Methods}
\subsection{Synthesis}
\subsubsection{Synthesis of \ce{Li2$M$N2} precursors}
The synthesis of \ce{Li2$M$N2} ($M$ = Zr, Hf) was adapted from prior reports.\cite{barker1974reactionsLi2HfN2} 
For \ce{Li2ZrN2}, \ce{Li3N} (Sigma Aldrich, $\geq$99.5\%, 60 mesh) and \ce{ZrN} (Alfa Aesar, 99\%) were combined (ca.\ 1.3 : 1 mole ratio, ca.\ 100\% excess \ce{Li3N}), homogenized with an agate mortar and pestle, loaded into a Zr crucible with a Zr lid, placed in an open quartz ampule, loaded in a quartz process tube, and transferred from the glovebox to the furnace without air exposure. The samples were heated under flowing nitrogen (60 sccm, 99.999\% purity) at +10 °C/min to 900 °C or 1000 °C, dwelled for between 3 and 12 h, cooled naturally, and recovered into the glovebox. Excess \ce{Li3N} volatilized away from the sample and reacted with the sacrificial quartz ampule. Regrinding  (sometimes with small additions of \ce{Li3N}) and reheating were necessary to achieve phase purity, as \ce{ZrN} and an unidentified (but presumably Li-rich) phase were often observed owing to variable \ce{Li3N} evaporation rates. 
For \ce{Li2HfN2}, a similar method was used, but Hf powder (Goodfellow, 95\%) was used as a precursor rather than the binary nitride. 

\subsubsection{Synthesis of \ce{Mg$M$N2}}
Samples were prepared by mixing \ce{Li2$M$N2 + Mg$X_2$} ($X$ = Cl, Br), homogenizing in an agate mortar and pestle, compressing ca.\ 100~mg pellets (0.25 inch diameter) at approximately 1 ton pressure, and heating. Anhydrous halides were used: \ce{MgCl2} (Sigma-Aldrich, 99.99\%, AnhydroBeads) and \ce{MgBr2} (Thermofisher, 98\%, anhydrous). Samples were sealed in quartz ampules under vacuum ($< 30$~mTorr) and heated in muffle furnaces at temperatures and times specified in the text.
Ramp rates were either +5°C/min or +10°C/min, and samples were allowed to cool naturally in the furnace. Samples were recovered into an argon glovebox for initial PXRD. To remove the halide byproduct, samples were removed from the glovebox and washed three times with water and once with isopropanol. The resulting powders were generally brick red to dark grey, depending on synthesis conditions. 

\subsection{Characterization}
\subsubsection{\textit{Ex situ} PXRD}
The products of all reactions were characterized by powder X-ray diffraction (PXRD). Laboratory X-ray diffraction patterns were collected on a Rigaku Ultima IV diffractometer with Cu K$\alpha$ X-ray radiation at room temperature. All samples were initially prepared for PXRD measurements inside the glovebox; powder was placed on off-axis cut silicon single crystal wafers to reduce background scattering and then covered with polyimide tape to impede exposure to atmosphere. After \ce{MgZrN2} was determined to be moderately air stable, PXRD patterns were collected without polyimide tape to decrease the background signal.

\subsubsection{\textit{In situ} SPXRD}
\textit{In situ} sychrotron powder X-ray diffraction (SPXRD) experiments were conducted at beamline 2-1 of the Stanford Synchrotron Radation Lightsource (SSRL) with a 17~keV X-ray energy ($\lambda = 0.7293$ \AA{}). 
Reactant mixtures were loaded into 0.5 mm OD (0.01 mm wall thickness) quartz capillaries and flame-sealed under vacuum.
These capillaries were subsequently nested inside 1.0 mm OD quartz capillaries (0.01 mm wall thickness) and mounted in an Anton Paar HTK 1200N heating stage. 
The capillary was rotated with a frequency of 1 Hz during the experiments. 
Diffraction patterns were collected with a small area detector (Pilatus 100K) at a 700 mm sample-to-detector distance.
Each 1D pattern was stitched together from 15 separate 2D exposures (1~s each) at 2° steps between 6° and 36° $2\theta$, with radial integration and merging conducted on-the-fly with custom python code.
One full pattern was collected every 50~s, accounting for motor movements.

\subsubsection{Analysis of diffraction data}
\textit{Ex situ} PXRD patterns were analyzed using TOPAS Professional v6.\cite{coelho2018topas} 
To create the \ce{MgZrN2} and \ce{MgHfN2} structures in space group $R\overline{3}m$, $\alpha$-\ce{NaFeO2} was used as a starting point (ICSD Collection Code 187705). 
Na was replaced with Mg, Fe was replaced with Zr (or Hf), and O was replaced with N.
Subsequently, this model was refined against the data by first fitting lattice parameters, then crystallite size broadening (Lorentzian), then displacement parameters ($B_{iso}$). 
The background was modeled with a 10-term polynomial.
As peak intensities and peak shapes were poorly captured by the $R\overline{3}m$ phase alone, rocksalt \ce{Mg$M$N2} ($Fm\overline{3}m$) was also added to the model, and a similar fitting sequence was followed.\cite{rom2021bulk}
In the case of \ce{MgHfN2}, the (003) reflection of the $R\overline{3}m$ phase was under-fit unless a preferred orientation term was used. We therefore modeled preferred orie

Sequential Rietveld refinements were conducted onri \textit{in situ} SXPRD datasets using TOPAS Professional v6.\cite{coelho2018topas} Lattice parameters, background terms, and scale factors were refined for each phase as a function of temperature, while atomic coordinates and occupancies were held constant at the initial values of the reference structure. A weighted scale factor (W.S.F.) $Q$ was calculated for each phase $p$ as a product of scale factor $S$, cell volume $V$, and cell mass $M$: $Q_p = S_p \bullet V_p \bullet W_p$. 
We note that amorphous and liquid phases are inherently not observed in powder diffraction measurements and therefore cannot be accurately included in this analysis. 
A Lorentzian size broadening term was refined for each phase to model the peak shape using the pattern showing the greatest intensity of the relevant phase; this term was then fixed for the sequential refinements to better account for changes in intensity. 
To help stabilize the sequential refinement, isotropic displacement parameters ($B_\mathrm{iso}$) were fixed at 1~\AA{}$^2$ for all atoms, but we note that this is likely not physical for a variable temperature investigation.

\subsubsection{Optical measurements}
UV-vis measurements were conducted on a Cary 6000 UV-Vis-NIR spectrometer. 
\ce{BaSO4} was used as a white reflectance standard. 
Absorbance was calculated with the Kubelka-Munk transformation, $k/s = (1-R)2 / 2R$ where $R$ is the reflectance, $k$ is the apparent absorption coefficient, and $s$ is the apparent scattering coefficient. 

\subsection{Computational methods}
The total energies of ternary nitrides and their decomposition products were calculated using density functional theory (DFT) \cite{DFT} as implemented in the Vienna ab initio Simulation Package (VASP).\cite{VASP} The formation enthalpy for each compound was computed using Fitted Elemental-phase Reference Energies (FERE) to correct the chemical potentials.\cite{FERE} These calculations were performed using the generalized gradient approximation (GGA) with the Perdew-Burke-Ernzerhof (PBE) exchange-correlation functional.\cite{GGA} Atomic cores were modeled using projector-augmented-wave (PAW) pseudopotentials.\cite{PAW} The pseudopotentials were those used in the FERE standard. To converge the total energy with the relatively hard nitrogen pseudopotential used in this standard, a plane-wave cutoff energy of 450 eV was used. The \textbf{k}-point grids were generated automatically with 20 subdivisions along each reciprocal lattice vector. Spin degrees of freedom were treated explicitly for Fe and Cu atoms, testing a ferromagnetic configuration as well as several antiferromagnetic configurations in the $\alpha$-NaFeO$_2$ structure, and using the lowest energy spin configurations when calculating reaction energies. For each composition, structure, and spin configuration, the atomic positions and the shape and volume of the primitive cell were optimized using the conjugate gradient algorithm. Repeated relaxations of cell shape and volume were performed for numerical reasons. Each relaxation is considered converged when the difference in total energy between steps is less than $10^{-6}$ eV. Gibbs free energies of formation as a function of temperature are calculated using a descriptor for the vibrational entropy contribution developed by Bartel \textit{et al.}\cite{bartel_physical_2018}

\section{Acknowledgements}
This work was authored at the National Renewable Energy Laboratory, operated by Alliance for Sustainable Energy, LLC, for the U.S. Department of Energy (DOE) under Contract No. DE-AC36-08GO28308. 
Primary funding provided by DOE Basic Energy Sciences Early Career Award ``Kinetic Synthesis of Metastable Nitrides.''
M.Q.P. acknowledges support from the U.S. Department of Energy, Office of Science, Office of Workforce Development for Teachers and Scientists (WDTS) under the Science Undergraduate Laboratory Internships Program (SULI).
Calculations were performed using National Renewable Energy Laboratory computational resources. 
M.J. and V.S. acknowledge support from the National Science Foundation (Grant No.\ DMR-1945010) for the computational portion of this study.
C.E.R. and J.R.N. acknowledge support from the National Science Foundation (Grant No.\ DMR-2210780).  
The views expressed in the article do not necessarily represent the views of the DOE or the U.S. Government. The U.S. Government retains and the publisher, by accepting the article for publication, acknowledges that the U.S. Government retains a nonexclusive, paid-up, irrevocable, worldwide license to publish or reproduce the published form of this work, or allow others to do so, for U.S. Government purposes. % or add a cover sheet: https://thesource.nrel.gov/publishing/disclaimers 
Use of the Stanford Synchrotron Radiation Lightsource, SLAC National Accelerator Laboratory, is supported by the U.S. Department of Energy, Office of Science, Office of Basic Energy Sciences under Contract No. DE-AC02-76SF00515. 
Thanks to Stephan Lany for the GW calculated absorbance spectrum.

\section{Supportin Information}
The supporting information contains additional references.\cite{schon2000investigation}

\bibliography{main}

\clearpage

\section{Supporting Information for: 
Ion exchange synthesizes layered polymorphs of \ce{MgZrN2} and \ce{MgHfN2}, two metastable semiconductors
}

\renewcommand{\thefigure}{S\arabic{figure}}
\setcounter{figure}{0}
\renewcommand{\thetable}{S\arabic{table}}
\setcounter{table}{0}

\tableofcontents

\addcontentsline{toc}{section}{Additional \textit{ex situ} XRD experiments on \ce{MgZrN2}}
\section{Additional XRD experiments and structural analysis on \ce{MgZrN2}}
The \ce{MgZrN2} samples did not exhibit obvious signs of degradation after exposure to air and water. Figure S1 shows patterns for the as-synthesized \ce{Li2ZrN2 + MgBr2 -> MgZrN2 + LiBr} (protected from air and moisture with polyimide tape) compared to the XRD pattern of the same sample after washing with water. In the as-synthesized sample, the large background scattering of the polyimide tape obscures the (003) reflection of the \ce{MgZrN2} phase. Washing clearly removes the LiBr phase, and does not shift the peaks of the \ce{MgZrN2} phase. As the washed sample was measured in air (without polyimide tape), the (003) reflection appears prominently.

\begin{figure}[ht!]
    \centering
    \includegraphics[width=0.7\linewidth]{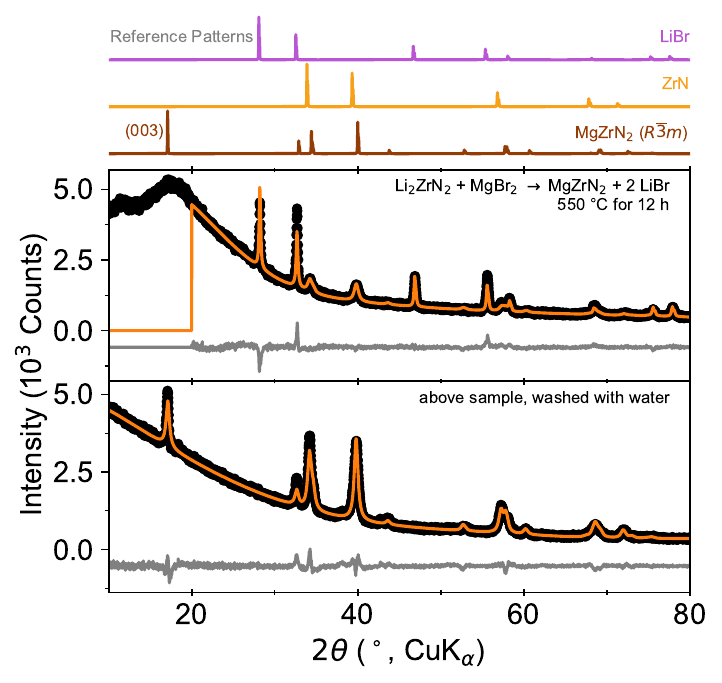}
    \caption{PXRD of the reaction product between \ce{Li2ZrN2 + MgBr2} heated at 550 °C for 12 h, compared with the powder after washing with water. 
    % For the unwashed sample (top) the powder is protected from atmosphere with a polyimide tape, which produces a large background below 30 ° $2\theta$. The washed sample was measured without this tape.
    }
    \label{fig:xrd_washed_v_unwashed}
    
\end{figure}

\begin{figure}[ht!]
    \centering
    \includegraphics[width=0.6\linewidth]{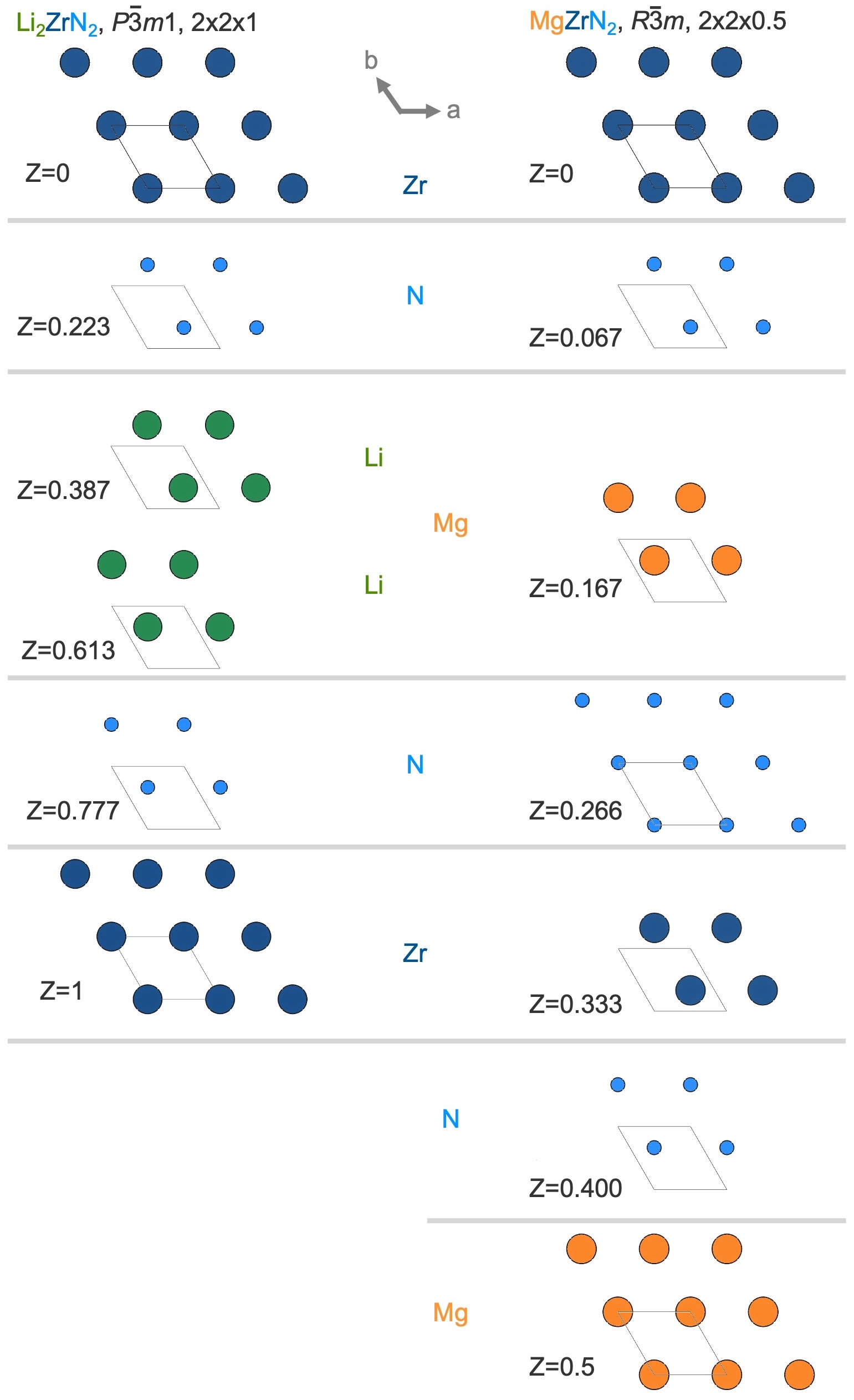}
    \caption{Visual comparison of the stacking sequence for \ce{Li2ZrN2} (left, 2x2x1 supercell) and \ce{MgZrN2} (right, 2x2x0.5 supercell).}
    \label{fig:structural_relationships}
\end{figure}
This ion exchange synthesis is topotactic. The Zr-layers are retained from the \ce{Li2ZrN2} precursor to the \ce{MgZrN2} product in the $\alpha$-\ce{NaFeO2} structure type. However, this transformation involves a shift of the Zr-containing layers and the anion sublattice. Figure S2 shows that \ce{Li2ZrN2} exhibits hexagonal-close-packing for the anion sublattice and cubic-close-packing for the cation sublattice, with Li in the tetrahedral voids between the anion layers. The cation sublattice of \ce{MgZrN2} is also cubic-close-packed, but the anion sublattice shifts from hexagonal- to cubic-close-packed.

\begin{figure}[ht!]
    \centering
    \includegraphics[width=0.7\linewidth]{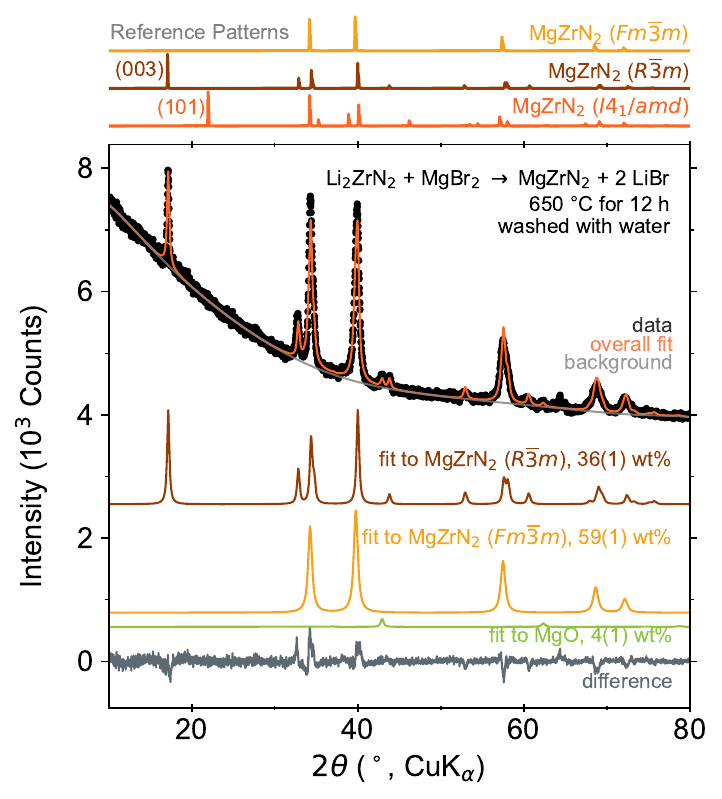}
    \caption{PXRD of the washed reaction products for \ce{Li2ZrN2 + MgBr2} heated at 650 °C for 12 h. % in a sealed ampule under vacuum.
    }
    \label{fig:xrd_MgZrN2_89B}
\end{figure}

\begin{table}[ht!]
\caption{Summary of weight percents of reaction products between \ce{Li2ZrN2 + MgBr2} heated to 650 °C for different dwell times. Phase fractions determined by Rietveld analysis, with fits shown in the associated figures.}
\begin{tabular}{lcccc}
Dwell time & \ce{MgZrN2} ($R\overline{3}m$) & \ce{MgZrN2} ($Fm\overline{3}m$) & MgO  & Fit \\ \hline
10 min     & 47(1) wt\%         & 53(1) wt\%          & -    & Figure \ref{fig:xrd_MgZrN2_main}      \\
12 h       & 36(1) wt\%         & 59(1) wt\%          & 4(1) wt\% & Figure S3     \\ \hline
\end{tabular}
\label{tab:rr_phaseFrac}
\end{table}

Reaction temperature and dwell time influences relative amounts of $R\overline{3}m$ and $Fm\overline{3}m$ \ce{MgZrN2} present in the sample. Higher temperatures and longer heating times tend to increase the fraction of the cation-disordered phase ($Fm\overline{3}m$). For example, identical reaction mixtures heated to 650 °C for a 10 min dwell time (Figure \ref{fig:xrd_MgZrN2_main}) and a 12 h dwell time (Figure S3) showed different phase fractions (Table S1). These differing structures also lead to different optical properties (Figure S4). While we did not optimize the synthesis conditions for phase-pure $R\overline{3}m$ \ce{MgZrN2} here, this finding suggests that cooler reaction temperatures and shorter reactions times would be preferable.

\clearpage
\addcontentsline{toc}{section}{Additional UV-vis experiments on \ce{MgZrN2}}
\section{Additional UV-vis experiments on \ce{MgZrN2}}

The reaction products appear reddish, although this color becomes darker with higher reaction temperatures or longer reaction times. Figure S4 compares the spectra of two samples of \ce{MgZrN2} reacted at 650 °C for either 10 min or 12 h. The powder from the longer reaction (black trace) looks black and absorbs more light at low photon energy compared to the 10 min reaction. In contrast, the 10 min reaction shows a sharper absorption onset near 2 eV, consistent with the calculated absorption onset for $R\overline{3}m$ \ce{MgZrN2}. The 10 min reaction was also presented in the main text (Figure \ref{fig:uv_vis_main}). 

\begin{figure}[ht!]
    \centering
    \includegraphics[width=0.7\linewidth]{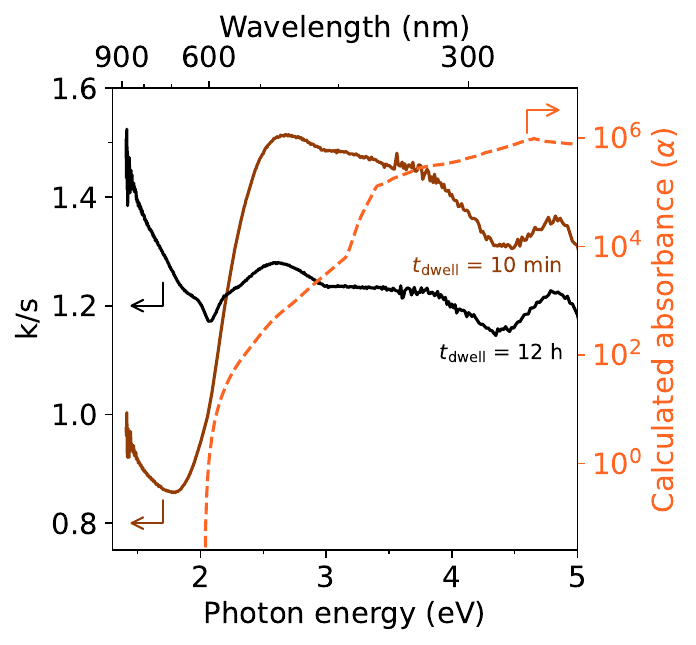}
    \caption{UV-vis spectra for the washed powders of \ce{MgZrN2} prepared from the reaction between \ce{Li2ZrN2 + MgBr2} heated to 650 °C for two different dwell times: 10 min (brown trace) and 12 h (black trace). For reference, the calculated absorbance is also shown.}
    \label{fig:uv_vis_suppInfo}
    %Source: http://127.0.0.1:8888/notebooks/Documents/02_Nitrides/UVvis/UV_vis_MgZrN2.ipynb 
\end{figure}

\clearpage

\addcontentsline{toc}{section}{\textit{In situ} SPXRD experiments}
\section{\textit{In situ} SPXRD experiments}
Figure S5 shows a larger version of the heatmap presented in Figure \ref{fig:inSitu_main}, along with select fits conducted as part of the Rietveld analysis.

\begin{figure}[ht!]
    \centering
    \includegraphics[width = \textwidth]{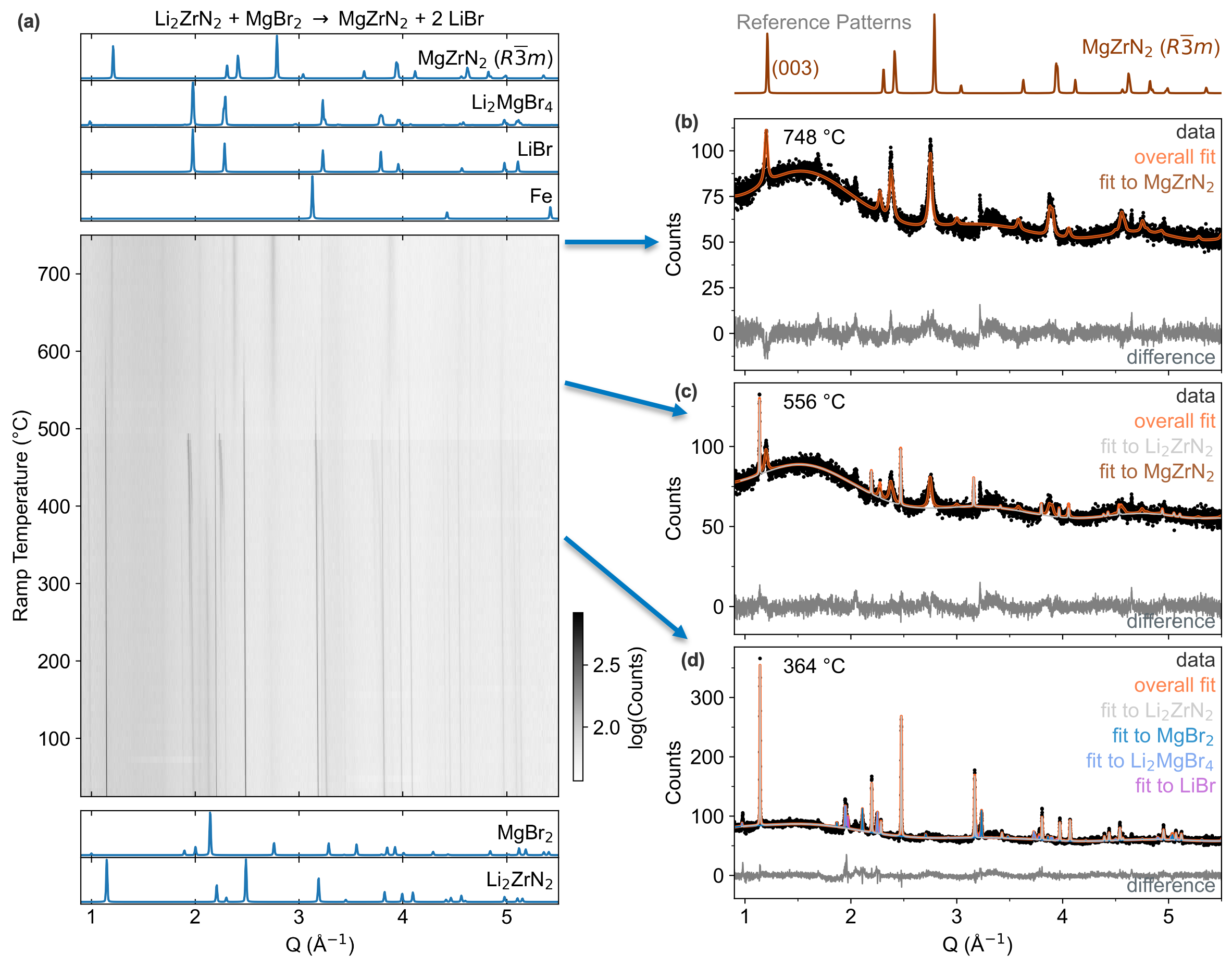}
    \caption{\textit{In situ} SPXRD of \ce{Li2ZrN2 + MgBr2} heated at +10°C/min showing (a) all patterns as a heatmap and select patterns at (b) 364 °C, (c) 556 °C, and (d) 748 °C fit via Rietveld analysis.}
    \label{fig:in_situ_Mg_supp}
\end{figure}

\textit{In situ} SPXRD experiments were conducted in attempts to synthesize layered $R\overline{3}m$ structures for \ce{FeZrN2} (Figure S6), \ce{CuZrN2} (Figure S7), and \ce{ZnZrN2} (Figure S8).
Strong reflections are observed for metallic Fe, Cu, and Zn, along with rocksalt ZrN, beginning as low as 100~°C in the case of Fe. 
These reduced metals indicate the loss of nitrogen gas from the structure (\ce{3 $A^{2+}$ + 2 N^{3-} -> 3 $A$ + N2}), and the rapid decomposition of any \ce{$A$ZrN2} ($A$ = Fe, Cu, Zn) that may be forming. 
This gas release caused the Fe-containing capillary to pop at approximately 400 °C, as seen by the sharp change in the diffraction. 
The loss of nitrogen from the solid suggests that \ce{FeZrN2}, \ce{CuZrN2}, and \ce{ZnZrN2} are not stable at the temperatures necessary for solid-state diffusion between \ce{Li2ZrN2} and the precursor salts ($\geq$200~°C).
Alternatively, self-heating from the exothermic ion exchange reaction may drive local temperatures above the decomposition point of the ternary nitrides. 

We attempted to mitigate self-heating for the Zn-based reaction via the addition of a LiCl/KCl eutectic flux (50 wt\%), but decomposition was still observed (Figure S8).
Metallic \ce{Zn} appears in the diffraction patterns near 200 °C, indicating decomposition.
There are small peaks that are we were unable to index, including a small peak near Q = 1.2~\AA{}$^{-1}$ that aligns well with the supercell reflections for the possible \ce{ZnZrN2} polymorphs: (003) for the $R\overline{3}m$ structure, or (001) for the $P3m1$ structure (Figure S8b-d).
However, our attempts to fit these patterns with these phases were unsuccessful. 
Interestingly, ball-milling the \ce{Li2ZrN2 + ZnBr2} with the LiCl/KCl flux resulted in apparent reactivity between the KCl and the \ce{ZnBr2}, as the \ce{ZnBr2} phase is not visible in the initial diffraction pattern (Figure S8d). 
Instead, broad peaks that index to a rocksalt phase--which we fit as K(Cl,Br) are present. 
This phase likely contains the Zn, but we were unable to refine the metal site occupancy owing to the similar scattering powers of K and Zn.
Reactivity between \ce{ZnBr2} and \ce{KCl} is not to blame for the formation of Zn metal; \textit{ex situ} experiments without a LiCl/KCl flux also yielded metallic Zn as a decomposition product, even with low reaction temperatures (Figures S9, S10).
Further \textit{in situ} SPXRD experiments with higher energy and higher flux X-rays are needed to better understand this reaction process.

\begin{figure}[ht!]
    \centering
    \includegraphics[width = \textwidth]{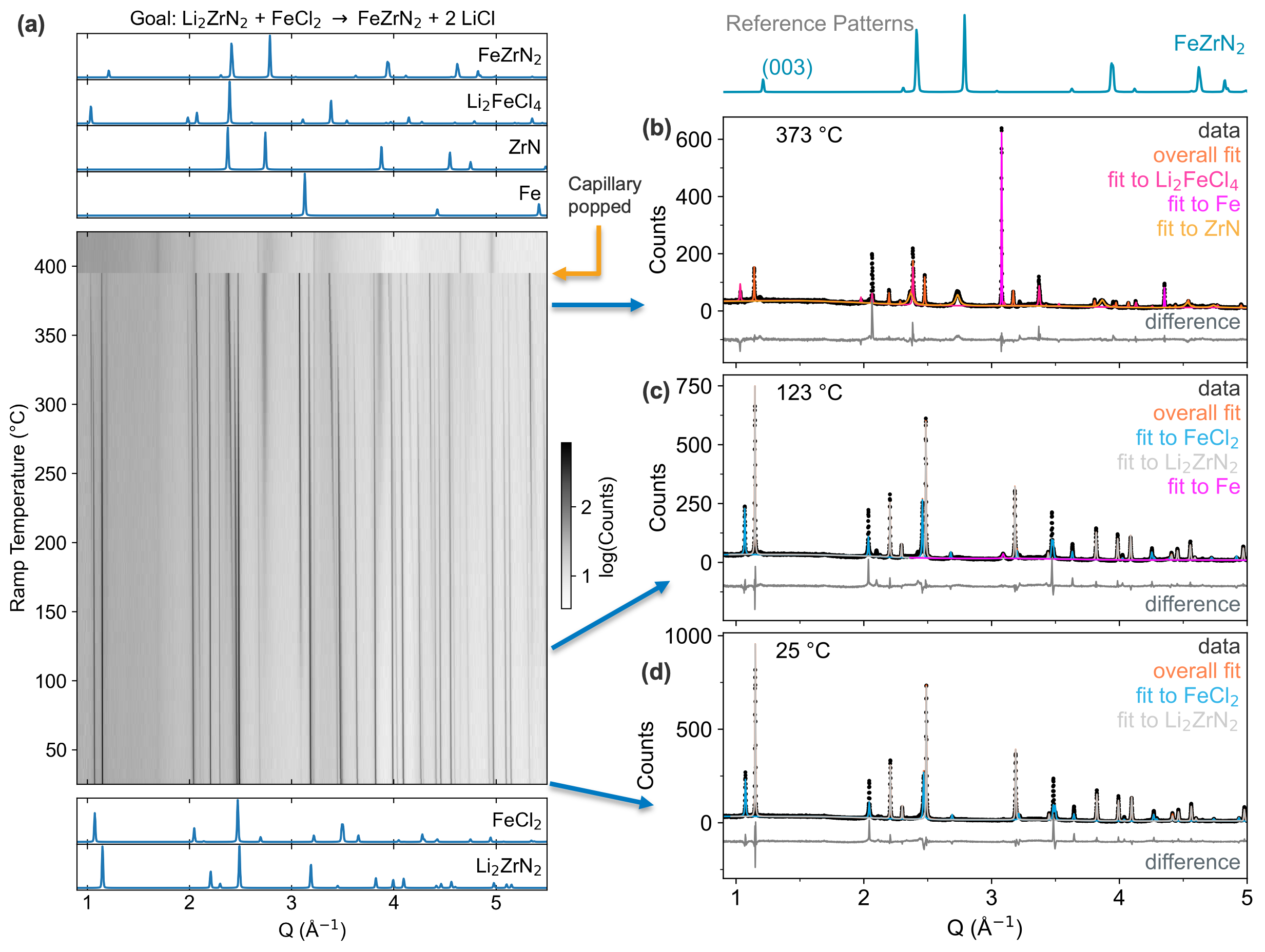}
    \caption{\textit{In situ} SPXRD of \ce{Li2ZrN2 + FeCl2} heated at +10°C/min showing (a) all patterns as a heatmap and select patterns at (b) 373 °C, (c) 123 °C, and (d) 25 °C fit via Rietveld analysis. The \ce{FeZrN2} reference pattern was generated with VESTA by replacing Mg in the $R\overline{3}m$ \ce{MgZrN2} structure with Fe.}
    \label{fig:in_situ_Fe}
\end{figure}

\begin{figure}[ht!]
    \centering
    \includegraphics[width = \textwidth]{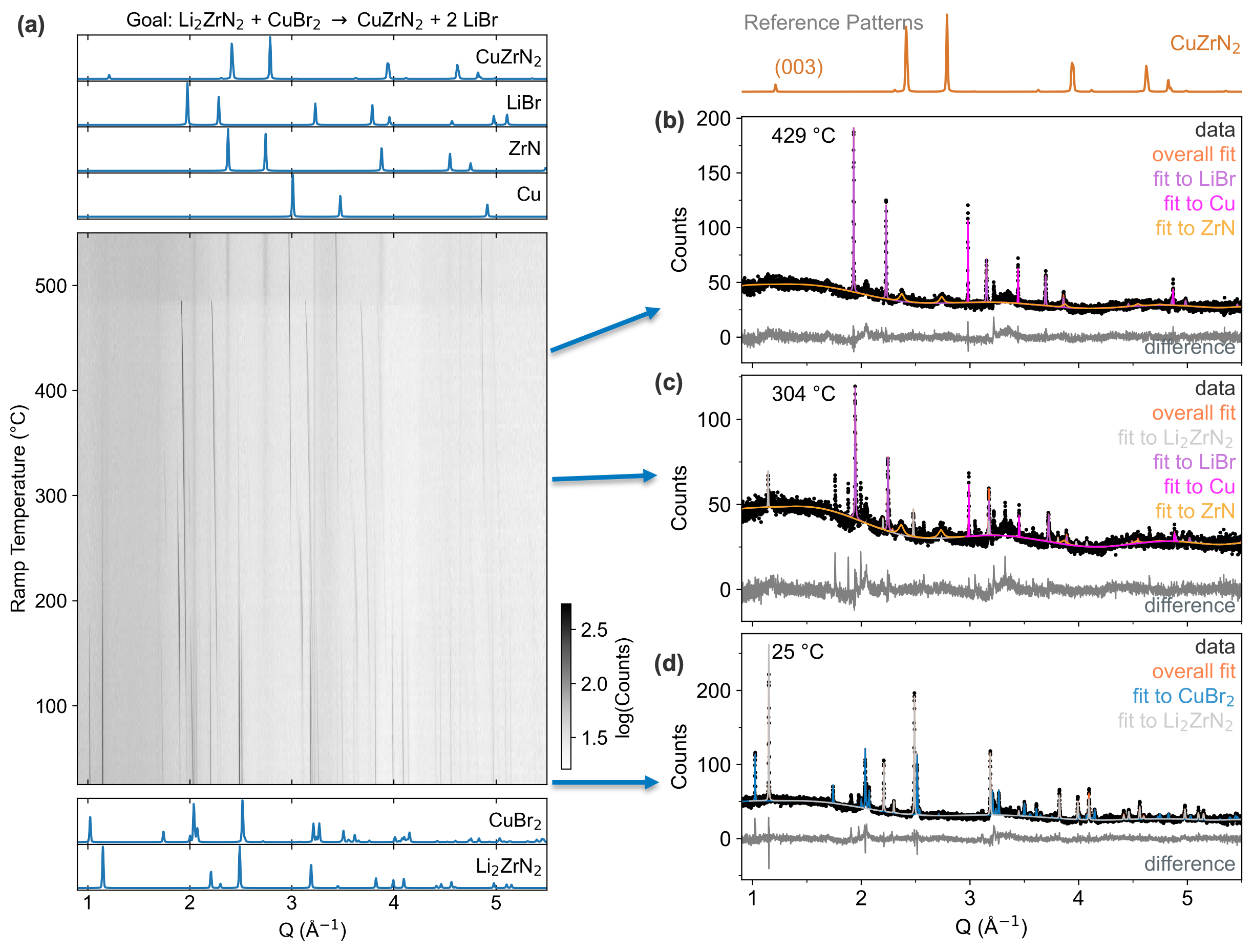}
    \caption{\textit{In situ} SPXRD of \ce{Li2ZrN2 + CuBr2} heated at +10°C/min showing (a) all patterns as a heatmap and select patterns at (b) 429 °C, (c) 304 °C, and (d) 25 °C fit via Rietveld analysis. The \ce{CuZrN2} reference pattern was generated with VESTA by replacing Mg in the $R\overline{3}m$ \ce{MgZrN2} structure with Cu.}
    \label{fig:in_situ_Cu}
\end{figure}

\begin{figure}[ht!]
    \centering
    \includegraphics[width = \textwidth]{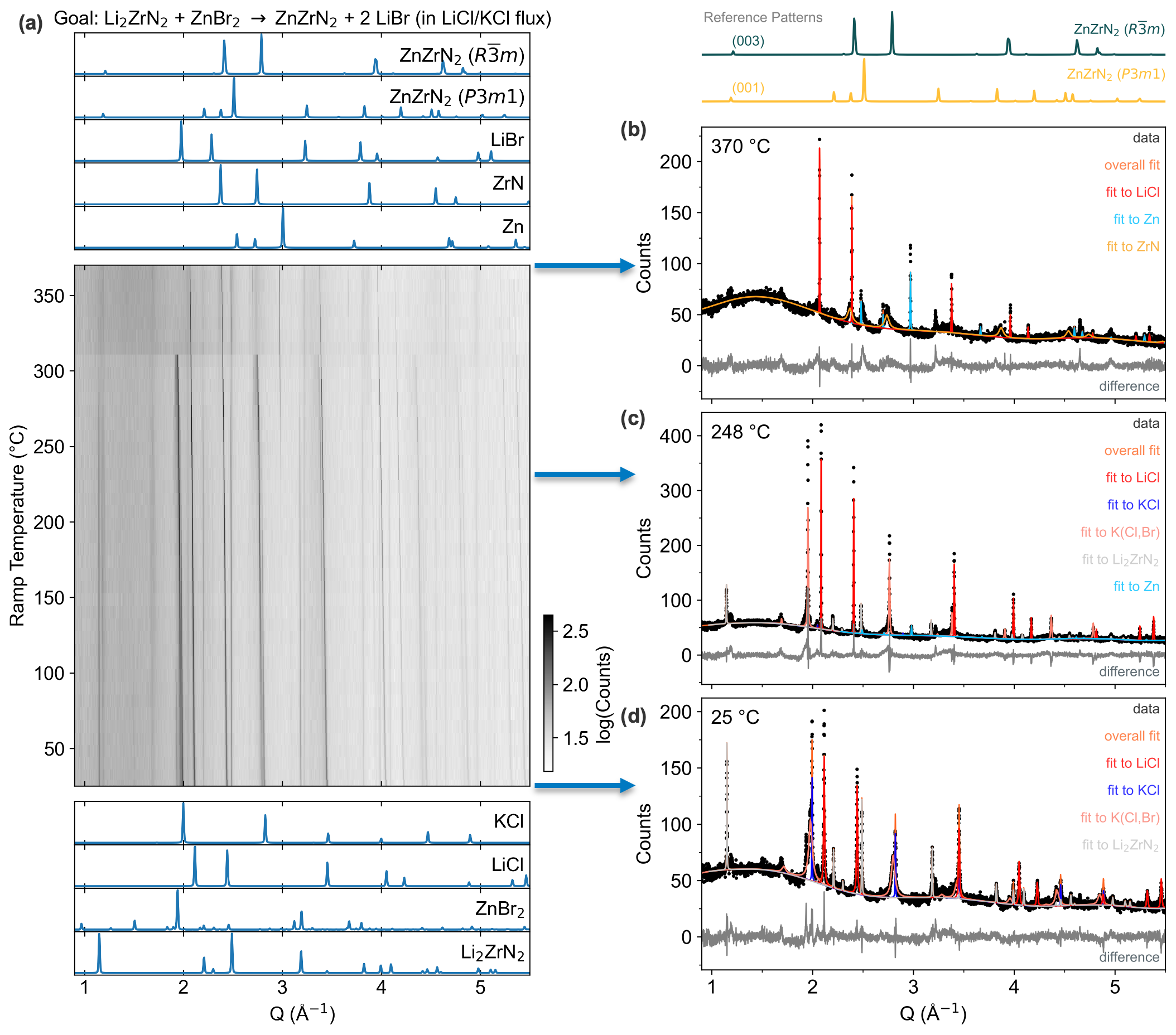}
    \caption{\textit{In situ} SPXRD of \ce{Li2ZrN2 + ZnBr2} (ball-milled with LiCl/KCl eutectic as a heat sink) heated at +10°C/min showing (a) all patterns as a heatmap and select patterns at (b) 370 °C, (c) 248 °C, and (d) 25 °C fit via Rietveld analysis. The \ce{ZnZrN2} $R\overline{3}m$ reference pattern was generated with VESTA by replacing Mg in the $R\overline{3}m$ \ce{MgZrN2} structure with Fe. The $P3m1$ pattern was generated from mp-1014244.}
    \label{fig:in_situ_Zn}
\end{figure}

\clearpage
\addcontentsline{toc}{section}{\textit{Ex situ} synthetic attempts towards \ce{ZnZrN2}}
\section{\textit{Ex situ} synthetic attempts towards \ce{ZnZrN2}}

\textit{Ex situ} PXRD of reactions between \ce{Li2ZrN2 + ZnBr2} do not show evidence of \ce{ZnZrN2} (Figures S9 and S10). Instead, decomposition products are detected (\ce{ZnZrN2 -> Zn + ZrN + 1/6 N2}). The reaction held at 200 °C proceeded only partially (Figure S9), yielding \ce{ZrN + Zn} along with unreacted \ce{Li2ZrN2} and an intermediate \ce{Li2ZnBr4}. The \ce{Li2ZnBr4} forms via the combination of the \ce{LiBr} product and the \ce{ZnBr2} precursor. At 300 °C, the reaction proceeds completely within 10 h (Figure S10), with the PXRD showing only \ce{Zn + ZrN + 2 LiBr}. 

\begin{figure}[ht!]
    \centering
    \includegraphics[width=0.7\linewidth]{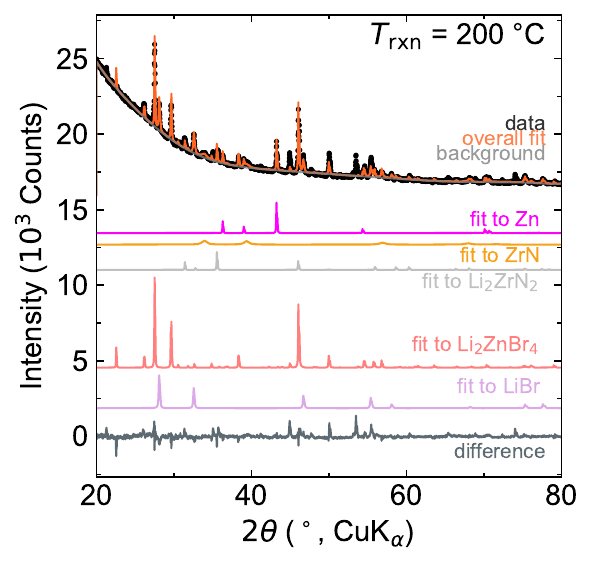}
    \caption{PXRD of the reaction products for \ce{Li2ZrN2 + ZnBr2} heated at 200 °C for 10 h.
    }
    \label{fig:xrd_ZnZrN2_44A}
\end{figure}

\begin{figure}[ht!]
    \centering
    \includegraphics[width=0.7\linewidth]{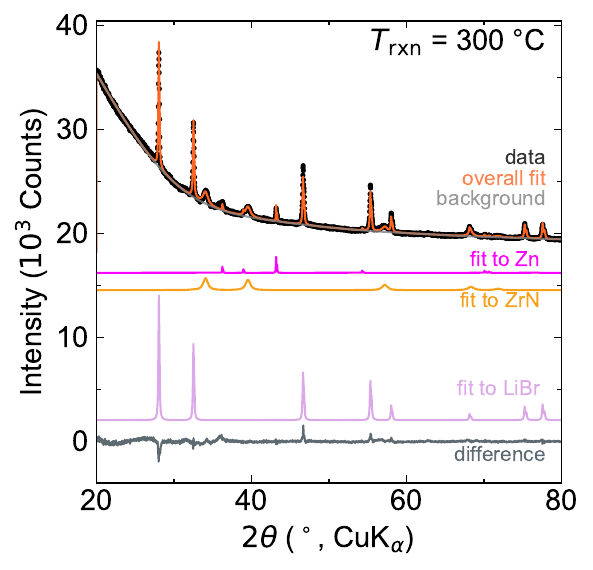}
    \caption{PXRD of the reaction products for \ce{Li2ZrN2 + ZnBr2} heated at 300 °C for 10 h.
    }
    \label{fig:xrd_ZnZrN2_44B}
\end{figure}

\clearpage
\addcontentsline{toc}{section}{Ion exchange thermodynamics}
\section{Ion exchange thermodynamics}
Temperature dependent free energy calculations for ion exchange reactions targeting $\alpha$-\ce{NaFeO2}-type \ce{ZnZrN2} and \ce{MgZrN2} are shown in Figure S11. 
While ion exchange targeting \ce{MgZrN2} is more exergonic than the reaction targeting \ce{ZnZrN2}, both reactions have a substantial forward driving force.
Therefore, the fact that \ce{MgZrN2} was synthesized while \ce{ZnZrN2} was not suggests that a different thermodynamic or kinetic factor matters.

\begin{figure}[ht!]
    \centering
    \includegraphics[width=0.7\linewidth]{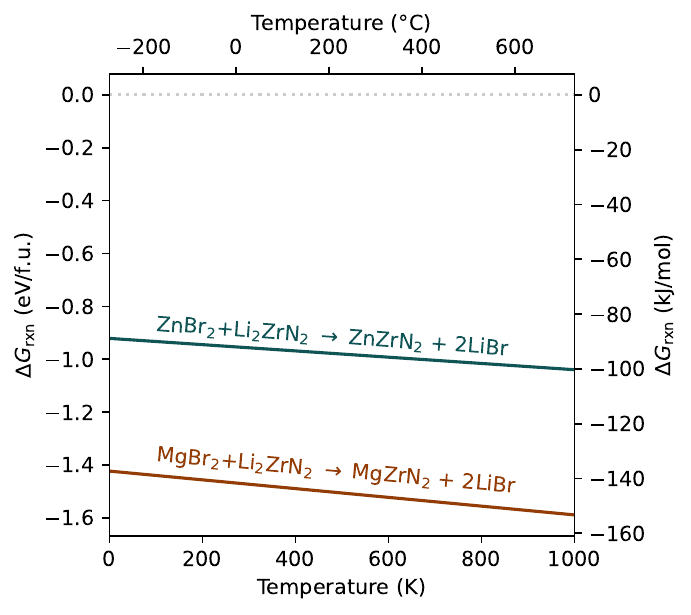}
    \caption{Overall $\Delta G_\mathrm{rxn}(T)$ energies for the reactions targeting \ce{ZnZrN2} and \ce{MgZrN2} beginning from bromide salts.}
    \label{fig:AZrN2_exchange_thermo}
%SourceL http://127.0.0.1:8889/notebooks/thermoFromMatt_MgZrN2.ipynb 
% /Users/crom/Documents/compmatscipy-master
% conda activate compMat (use a special environment)
% Navigate to that folder first, then activate the conda env
\end{figure}

Hess's law for reaction enthalpies allows us to consider the nitride and halide parts of the ion exchange reaction independently.
Consider the general reaction of interest here:
\begin{equation*}
\ce{Li2ZrN2 + $AX_2$ -> $A$ZrN2 + 2 Li$X$}
\end{equation*}
By Hess’s law,
\begin{equation*}
    \Delta H_\mathrm{rxn} = (\mathrm{products}) - (\mathrm{reactants}) = (2*\Delta H_{\mathrm{f, }\mathrm{Li}X} +  \Delta H_{\mathrm{f, }A\mathrm{ZrN_2}}) - (\Delta H_{\mathrm{f, }\mathrm{Li_2ZrN_2}} + \Delta H_{\mathrm{f, }AX_2})
\end{equation*}
For a given target compound \ce{AZrN2}, the components $\Delta H_\mathrm{Li_2ZrN_2}$ and $\Delta H_{AZrN_2}$ can be treated as a constant because they cannot be changed by experimental choices.
However, an experimenter can pick which halide $X$ to use for the reaction based on factors like melting point, solubility, and reaction enthalpy. 
To evaluate the thermodynamics of this choice, the overall Hess's law equation for the reaction can be simplified to
\begin{equation*}
    (2*\Delta H_{\mathrm{f, }\mathrm{Li}X}) - (\Delta H_{\mathrm{f, }AX_2})
\end{equation*}
Which is equivalent to the reaction \ce{2 Li + $AX$2 -> 2 Li$X$ + $A$}.
Generally, these values are substantially negative (Li$X$ halides are extremely stable), which means reactions tend to proceed forwards.
Figure S12 shows how choice of $X$ affects the ion exchange reaction energetics for the ion exchanges reactions presented here. 
Notably, reactions beginning from transition metal halides are substantially more exothermic than the \ce{Mg$X$2} ion exchanges, owing to the less negative $\Delta H_{\mathrm{f, }AX2}$ values for $A$ = Fe, Cu, Zn compared to Mg. 

\begin{figure}[ht!]
    \centering
    \includegraphics[width=0.7\linewidth]{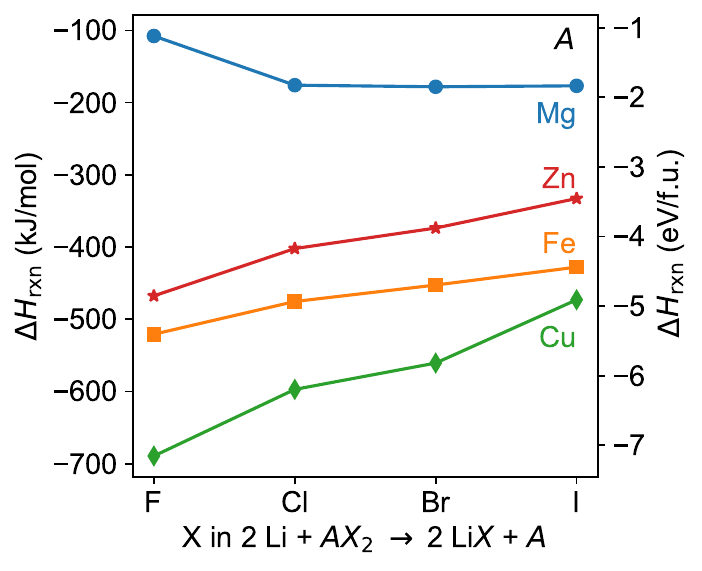}
    \caption{Reaction enthalpies calculated for \ce{2 Li + $AX$2 -> 2 Li$X$ + $A$} using $\Delta H_{\mathrm{f}}$ values from the CRC Handbook of Chemistry and Physics.}
    \label{fig:AX2_LiX_thermo}
    %Source: http://localhost:8889/notebooks/Documents/02_Nitrides/manuscript_layered_MgZrN2/Li2ZrN2_AZrN2_ion_Exchange_thermo.ipynb 
\end{figure}

In contrast, if the goal is just to consider stability of the \ce{AZrN2} phase, then overall reaction enthalpy is less relevant compared to the energy of the competing phases (e.g., elements and binaries). Therefore, we created Figure \ref{fig:dG_T} to show the relative stability of \ce{AZrN2} phases against decomposition. 

\clearpage
\addcontentsline{toc}{section}{Li-$M$-N entries in the PDF 2021 database}
\section{Li-$M$-N entries in the PDF 2021 database}
The data used for coloring the periodic table shown in Figure \ref{fig:ptable} were collected by searching the Powder Diffraction File (PDF) database using the following criteria:
\begin{itemize}
    \item Element filter = Yes for Li, N
    \item Element filter = Maybe for all metals and metalloids.
    \item Number of elements = 3
\end{itemize}
These data are presented in Table S2. 

\begin{longtable}{llll}%[ht!]
\caption{Li-$M$-N ternaries reported in the PDF 2021 database} 
% \begin{tabular}{llll}
\\ \hline
PDF \#      & Empirical formula     & Metal & SpaceGroup \\ \hline
00-007-0245 & Al Li3 N2             & Al    & 206        \\
01-074-0142 & Al Li3 N2             & Al    & 206        \\
01-086-3966 & Al Li3 N2             & Al    & 206        \\
03-065-3189 & Al Li3 N2             & Al    & 206        \\
01-074-1358 & B Li3 N2              & B     & 14         \\
01-075-3054 & B Li3 N2              & B     & 14         \\
00-040-1166 & B Li3 N2              & B     & 94         \\
01-075-3053 & B Li3 N2              & B     & 94         \\
01-080-2274 & B Li3 N2              & B     & 136        \\
01-075-3051 & B Li3 N2              & B     & 141        \\
01-075-3052 & B Li3 N2              & B     & 141        \\
00-013-0393 & B Li3 N2              & B     &            \\
00-016-0273 & B Li3 N2              & B     &            \\
00-050-1495 & B Li3 N2              & B     &            \\
00-023-1175 & Ba Li N               & Ba    &            \\
01-077-3911 & Ba2 Li N              & Ba    & 137        \\
01-077-3912 & Ba3 Li N              & Ba    & 194        \\
00-059-0159 & Ba3 Li N              & Ba    & 194        \\
01-086-1793 & Be Li N               & Be    & 14         \\
01-072-9363 & Ca Li N               & Ca    & 62         \\
00-018-0724 & Ca Li N               & Ca    &            \\
01-083-8888 & Ca3 Li2 N6            & Ca    & 51         \\
00-024-0603 & Ce Li2 N2             & Ce    & 164        \\
01-076-0449 & Ce Li2 N2             & Ce    & 164        \\
00-005-0605 & Co0.43 Li2.57 N       & Co    & 191        \\
03-065-9918 & Co0.43 Li2.57 N       & Co    & 191        \\
01-071-7305 & Co0.46 Li2.54 N       & Co    & 191        \\
01-073-3896 & Co0.53 Li1.99 N       & Co    & 191        \\
01-085-4405 & Co0.43 Li2.57 N       & Co    & 191        \\
00-036-0695 & Cr Li9 N5             & Cr    &            \\
01-089-2897 & Cr0.8 Li7.2 N4        & Cr    & 225        \\
01-079-1124 & Cr2 Li15 N9           & Cr    & 130        \\
01-074-3506 & Cu0.06 Li2.94 N       & Cu    & 191        \\
01-080-5162 & Cu0.167 Li2.833 N     & Cu    & 191        \\
01-080-5163 & Cu0.289 Li2.711 N     & Cu    & 191        \\
01-085-4412 & Cu0.3 Li2.7 N         & Cu    & 191        \\
03-065-9949 & Cu0.3 Li2.7 N         & Cu    & 191        \\
01-080-5164 & Cu0.348 Li2.652 N     & Cu    & 191        \\
01-070-3089 & Cu0.39 Li2.6 N        & Cu    & 191        \\
01-070-7316 & Cu0.404 Li2.576 N     & Cu    & 191        \\
01-070-7315 & Cu0.423 Li2.558 N     & Cu    & 191        \\
01-070-7313 & Cu0.428 Li2.572 N     & Cu    & 191        \\
01-072-6104 & Cu0.43 Li2.25 N       & Cu    & 191        \\
01-070-7314 & Cu0.43 Li2.57 N       & Cu    & 191        \\
01-074-3507 & Cu0.43 Li2.57 N       & Cu    & 191        \\
01-073-7690 & Cu2.81 Li0.1 N        & Cu    & 221        \\
01-073-7691 & Cu2.88 Li0.13 N       & Cu    & 221        \\
01-073-7695 & Cu2.94 Li0.85 N       & Cu    & 221        \\
01-073-7696 & Cu2.94 Li1.03 N       & Cu    & 221        \\
01-073-7692 & Cu3 Li0.42 N          & Cu    & 221        \\
01-073-7693 & Cu3 Li0.52 N          & Cu    & 221        \\
01-073-7694 & Cu3 Li0.64 N          & Cu    & 221        \\
01-080-0718 & Fe Li3 N2             & Fe    & 72         \\
00-020-0626 & Fe Li3 N2             & Fe    &            \\
01-080-0888 & Fe Li4 N2             & Fe    & 71         \\
01-083-6626 & Fe0.26 Li2.74 N       & Fe    & 191        \\
01-089-1762 & Fe0.63 Li2.37 N       & Fe    & 191        \\
01-073-9746 & Fe0.86 Li3 N2         & Fe    & 72         \\
00-055-0451 & Ga Li3 N2             & Ga    & 206        \\
01-074-0143 & Ga Li3 N2             & Ga    & 206        \\
01-077-6079 & Ga Li3 N2             & Ga    & 206        \\
03-065-3190 & Ga Li3 N2             & Ga    & 206        \\
00-028-0566 & Ge Li N               & Ge    &            \\
00-028-0565 & Ge Li2 N2             & Ge    &            \\
00-027-1240 & Ge Li8 N4             & Ge    & 206        \\
01-082-6921 & Ge0.67 Li3.33 N2      & Ge    & 206        \\
01-074-0161 & Ge10.67 Li53.33 N32   & Ge    & 206        \\
00-027-1239 & Ge2 Li N3             & Ge    & 36         \\
01-082-6922 & Ge2 Li N3             & Ge    & 36         \\
01-086-6831 & Ge2 Li N3             & Ge    & 36         \\
00-027-0293 & Hf Li2 N2             & Hf    & 164        \\
00-049-1181 & In Li3 N2             & In    &            \\
01-070-7485 & Li Mg N               & Mg    & 62         \\
01-079-5487 & Li Mg N               & Mg    & 62         \\
01-079-5488 & Li Mg N               & Mg    & 62         \\
01-079-5489 & Li Mg N               & Mg    & 62         \\
00-006-0702 & Li Mg N               & Mg    & 225        \\
01-072-1287 & Li Mg N               & Mg    & 225        \\
01-079-5490 & Li Mg N               & Mg    & 225        \\
01-079-5491 & Li0.24 Mg2.76 N1.838  & Mg    & 199        \\
01-079-5493 & Li0.24 Mg2.76 N1.92   & Mg    & 199        \\
01-079-5494 & Li0.24 Mg2.76 N1.92   & Mg    & 199        \\
01-079-5495 & Li0.24 Mg2.76 N1.92   & Mg    & 199        \\
01-079-5492 & Li0.48 Mg2.52 N1.84   & Mg    & 199        \\
01-070-7484 & Li0.51 Mg2.49 N1.83   & Mg    & 199        \\
01-079-5500 & Li1.09 Mg0.91 N0.97   & Mg    & 62         \\
01-079-5497 & Li1.10 Mg0.90 N0.96   & Mg    & 225        \\
01-079-5499 & Li1.10 Mg0.90 N0.96   & Mg    & 225        \\
01-079-5496 & Li1.11 Mg0.89 N0.96   & Mg    & 225        \\
01-079-5498 & Li1.11 Mg0.89 N0.96   & Mg    & 225        \\
01-070-7486 & Li1.12 Mg0.88 N0.96   & Mg    & 225        \\
01-089-1030 & Li0.66 Mn1.34 N       & Mn    & 136        \\
01-089-1031 & Li0.86 Mn1.14 N       & Mn    & 136        \\
01-089-1764 & Li2.27 Mn0.73 N       & Mn    & 191        \\
01-070-9584 & Li2.33 Mn0.67 N       & Mn    & 191        \\
01-070-9582 & Li24 Mn3 N10.86       & Mn    & 163        \\
01-070-9583 & Li6.23 Mn1.77 N3      & Mn    & 189        \\
01-074-9559 & Li7 Mn N4             & Mn    & 218        \\
00-051-1219 & Li7 Mn N4             & Mn    & 218        \\
01-070-9581 & Li7 Mn N4             & Mn    & 218        \\
01-089-4135 & Li7 Mn N4             & Mn    & 218        \\
01-081-0578 & Li Mo N2              & Mo    & 160        \\
00-047-1478 & Li Mo N2              & Mo    & 166        \\
01-079-1122 & Li6 Mo N4             & Mo    & 137        \\
01-070-6781 & Li N Na2              & Na    & 191        \\
01-070-6787 & Li N2 Na5             & Na    & 1          \\
01-070-6786 & Li N2 Na5             & Na    & 5          \\
01-070-6779 & Li2 N Na              & Na    & 191        \\
01-070-6780 & Li2 N2 Na4            & Na    & 129        \\
01-070-6783 & Li3 N2 Na3            & Na    & 6          \\
01-070-6782 & Li3 N2 Na3            & Na    & 115        \\
01-070-6778 & Li4 N2 Na2            & Na    & 65         \\
01-070-6776 & Li4 N2 Na2            & Na    & 139        \\
01-070-6777 & Li4 N2 Na2            & Na    & 225        \\
01-070-6785 & Li5 N2 Na             & Na    & 6          \\
01-070-6784 & Li5 N2 Na             & Na    & 123        \\
00-053-0436 & Li N4 Nb3             & Nb    & 193        \\
01-081-0211 & Li7 N4 Nb             & Nb    & 205        \\
01-078-6910 & Li N Ni               & Ni    & 187        \\
01-089-6875 & Li N Ni               & Ni    & 187        \\
01-072-6105 & Li1.35 N Ni0.79       & Ni    & 191        \\
01-086-5609 & Li2 N Ni0.67          & Ni    & 191        \\
01-086-5610 & Li2 N Ni0.67          & Ni    & 191        \\
01-074-3503 & Li2.15 N Ni0.85       & Ni    & 191        \\
01-071-9493 & Li2.57 N Ni0.43       & Ni    & 191        \\
01-074-3505 & Li2.63 N Ni0.37       & Ni    & 191        \\
01-074-3504 & Li2.93 N Ni0.07       & Ni    & 191        \\
01-089-6876 & Li5 N3 Ni3            & Ni    & 189        \\
01-086-2147 & Li5.69 N3 Ni2.31      & Ni    & 189        \\
01-086-5611 & Li5.69 N3 Ni2.31      & Ni    & 189        \\
03-065-3353 & Li7 N4 Ni             & Ni    & 218        \\
01-070-6922 & Li5 N4 Re             & Re    & 59         \\
01-071-6773 & Li5 N4 Re             & Re    & 59         \\
01-072-8199 & Li3 N2 Sc             & Sc    & 206        \\
00-026-1186 & Li N3 Si2             & Si    & 36         \\
00-050-0747 & Li N3 Si2             & Si    & 36         \\
01-070-3183 & Li N3 Si2             & Si    & 36         \\
01-076-0517 & Li N3 Si2             & Si    & 36         \\
01-086-6830 & Li N3 Si2             & Si    & 36         \\
00-040-1447 & Li18 N10 Si3          & Si    &            \\
01-078-8859 & Li2 N2 Si             & Si    & 61         \\
00-023-0365 & Li2 N2 Si             & Si    &            \\
00-040-1448 & Li21 N11 Si3          & Si    &            \\
00-007-0260 & Li5 N3 Si             & Si    & 206        \\
00-040-1446 & Li5 N3 Si             & Si    & 225        \\
01-074-0159 & Li53.33 N32 Si10.67   & Si    & 206        \\
01-083-4019 & Li54.88 N30.98 Si9.12 & Si    & 79         \\
00-040-1449 & Li8 N4 Si             & Si    &            \\
01-089-6184 & Li N Sr               & Sr    & 131        \\
00-023-0367 & Li N Sr               & Sr    &            \\
01-089-6183 & Li4 N2 Sr             & Sr    & 141        \\
01-081-0296 & Li0.977 N4 Ta3.023    & Ta    & 193        \\
01-081-0295 & Li0.983 N4 Ta3.017    & Ta    & 193        \\
01-081-0290 & Li1.8 N4 Ta2.2        & Ta    & 225        \\
01-081-0289 & Li1.85 N4 Ta2.15      & Ta    & 225        \\
01-071-3446 & Li4 N3 Ta             & Ta    & 73         \\
01-079-2303 & Li7 N4 Ta             & Ta    & 205        \\
01-078-7112 & Li8 N2 Te             & Te    & 109        \\
01-078-7113 & Li8 N2 Te             & Te    & 109        \\
00-025-0502 & Li2 N2 Th             & Th    & 147        \\
01-072-0794 & Li2 N2 Th             & Th    & 147        \\
00-006-0703 & Li5 N3 Ti             & Ti    & 206        \\
01-074-0160 & Li53.33 N32 Ti10.67   & Ti    & 206        \\
01-074-3519 & Li N2 U               & U     & 141        \\
00-025-0504 & Li2 N2 U              & U     & 147        \\
01-082-7245 & Li2 N2 U              & U     & 147        \\
01-072-7053 & Li7 N4 V              & V     & 137        \\
01-070-5088 & Li7 N4 V              & V     & 205        \\
01-072-7052 & Li7 N4 V              & V     & 205        \\
00-015-0216 & Li7 N4 V              & V     & 218        \\
01-072-7051 & Li7 N4 V              & V     & 218        \\
01-089-3703 & Li7 N4 V              & V     & 218        \\
00-049-1502 & Li N2 W               & W     & 146        \\
00-050-1482 & Li0.84 N2 W1.16       & W     & 194        \\
01-089-7046 & Li0.84 N2 W1.16       & W     & 194        \\
00-057-0399 & Li6 N4 W              & W     & 137        \\
01-079-1123 & Li6 N4 W              & W     & 137        \\
01-074-9122 & Li6 N4 W             & W     & 137        \\
01-072-1288 & Li N Zn               & Zn    & 216        \\
00-006-0467 & Li Zn N               & Zn    & 216        \\
01-080-8523 & Li N2 Zr              & Zr    & 164        \\
00-025-0506 & Li2 N2 Zr             & Zr    & 164        \\
00-050-0843 & Li2 N2 Zr             & Zr    & 164        \\
01-072-0793 & Li2 N2 Zr             & Zr    & 164        \\
01-089-0306 & Li2 N2 Zr             & Zr    & 164       \\ \hline
% \end{tabular}
\label{tab:PDF_entries}
\end{longtable}

The number of unique ternaries were estimated by counting the number of unique spacegroups reported for each metal $M$ (Table S3). 
We did not count PDF entries for which a spacegroup was not reported.
If there are multiple competing reports for the same phase (e.g., experimental ambiguity in structural determination), then our counts may be overestimates.
Additionally, the count of 10 unique structures for $M$ = Na struck us as an outlier.
These 10 unique structures originated from a single computational study.\cite{schon2000investigation} 
As these phases have not been experimentally determined, we marked the count for Li-Na-N ternaries as 0 in Figure \ref{fig:ptable}.
Other computationally-derived structures may be present in Table S2 that we have not identified.

\begin{table}[ht!]
\caption{Summary of unique space groups for Li-$M$-N ternaries shown in Table S2. There are 63 unique Li-$M$-N phases reported (* excluding Na, which are computational rather than experimental structures).}
\begin{tabular}{lc}
Metal & Unique space groups \\ \hline
Al    & 1                   \\
B     & 4                   \\
Ba    & 2                   \\
Be    & 1                   \\
Ca    & 2                   \\
Ce    & 1                   \\
Co    & 1                   \\
Cr    & 2                   \\
Cu    & 2                   \\
Fe    & 3                   \\
Ga    & 1                   \\
Ge    & 2                   \\
Hf    & 1                   \\
Mg    & 3                   \\
Mn    & 3                   \\
Mo    & 3                   \\
Na    & 10*                  \\
Nb    & 2                   \\
Ni    & 4                   \\
Re    & 1                   \\
Sc    & 1                   \\
Si    & 5                   \\
Sr    & 2                   \\
Ta    & 4                   \\
Th    & 1                   \\
Ti    & 1                   \\
U     & 2                   \\
V     & 3                   \\
W     & 3                   \\
Zn    & 1                   \\
Zr    & 1                  \\ \hline
% Total & 63 \\ \hline
\end{tabular}
\label{tab:unique_LiMN}
\end{table}

\end{document}